\documentclass[aps,pra,10pt,showpacs,twocolumn,superscriptaddress,longbibliography]{revtex4-1}
\usepackage{graphicx}
\usepackage[usenames]{color}
\usepackage{amssymb,amsmath}
\usepackage{siunitx}
\usepackage{xcolor}
\usepackage{setspace} 
\usepackage{float} 
\usepackage{tabularx}
\usepackage[linkcolor=blue,citecolor=blue,colorlinks=true,urlcolor=blue]{hyperref}
\newcommand{\eq}[1]{Eq.~(\ref{#1})}
\newcommand{\fig}[1]{Fig.~\ref{#1}}
\newcommand{\be}[1]{\begin{equation}\label{#1}}
\newcommand{\ee}{\end{equation}}
\usepackage{amsmath,esint}
\usepackage{comment}
\usepackage{lipsum}
\usepackage[toc]{appendix}

\begin{document}

\title{A general model and toolkit for the ionization of three or more electrons in strongly driven atoms using an effective Coulomb potential for the interaction between bound electrons}

\author{M. B. Peters}
\affiliation{Department of Physics and Astronomy, University College London, Gower Street, London WC1E 6BT, United Kingdom}
\author{G. P. Katsoulis}
\affiliation{Department of Physics and Astronomy, University College London, Gower Street, London WC1E 6BT, United Kingdom}
\author{A. Emmanouilidou}
\affiliation{Department of Physics and Astronomy, University College London, Gower Street, London WC1E 6BT, United Kingdom}

\begin{abstract}
We formulate a three-dimensional semi-classical model to address  triple and double ionization in three-electron  atoms driven by  intense infrared laser pulses. During time propagation, our model fully accounts for the Coulomb singularities, the magnetic field of the laser pulse and for the motion of the nucleus  at the same time as for the motion of the three electrons. The framework we develop is general and  can account  for multi-electron ionization in strongly-driven atoms with more than three electrons. To avoid unphysical autoionization arising in classical models of three or more electrons, we replace the Coulomb potential between pairs of bound electrons with effective Coulomb potentials. The Coulomb forces between  electrons that are not both bound are fully accounted for. We develop a  set of criteria to determine when electrons become bound during  time propagation. We compare  ionization spectra obtained with the model developed here and with the Heisenberg  model that includes a potential term  restricting an electron   from closely approaching the core. Such spectra include the sum of the electron momenta along the direction of the laser field as well as the correlated electron momenta. We also compare these results   with experimental ones. \end{abstract}

\date{\today}

\maketitle

\section{Introduction}
 Multi-electron ionization in atoms and molecules driven by intense infrared laser fields is of fundamental interest since it is mediated by electronic correlation. 
The theoretical study of correlated multi-electron  dynamics in strongly-driven atoms and molecules poses a significant challenge. Indeed, three-dimensional (3D) ab-initio quantum-mechanical methods are mostly limited to double ionization in two-electron atoms \cite{Parker_2000,PhysRevLett.96.133001}. Various quantum mechanical \cite{PhysRevA.93.023406,PhysRevA.103.053123} and semi-classical techniques \cite{PhysRevA.63.043416,PhysRevA.65.021406,PhysRevA.78.023411} that include the Coulomb singularity have  been developed to address double ionization. However, for three-electron escape,  due to the larger degree of complexity involved, only few theoretical studies exist that have a number of approximations. These studies include classical models with reduced-dimensionality  \cite{PhysRevA.64.053401}
  and with soft core Coulomb potentials \cite{PhysRevA.78.013401,PhysRevLett.97.083001,Zhou:10,Tang:13,Ho:07}, reduced-dimensionality quantum mechanical treatments   \cite{PhysRevA.98.031401,Prauzner_Bechcicki_2021,Efimov:21} and semi-classical models with Heisenberg potentials \cite{PhysRevA.104.023113}. On the experimental front, several studies have  addressed multi-electron ionization in  strongly-driven Ar and Ne \cite{PhysRevLett.84.447,Rudenko_2008,PhysRevA.86.043402,Herrwerth_2008,Zrost_2006,PhysRevLett.93.253001,SHIMADA2005221}.  For weak fields, striking angular patterns of  three-electron escape and the underlying  collision mechanisms were identified with 3D semi-classical models and ab-initio quantum mechanical techniques \cite{Emmanouilidou_2006,PhysRevLett.100.063002,PhysRevLett.108.053001,PhysRevLett.110.063001}. 
 
The main challenge facing quantum mechanical studies of triple ionization in strongly-driven systems  is the significant amount of   computational resources. This explains  the development of reduced dimensionality quantum mechanical models \cite{PhysRevA.98.031401,Prauzner_Bechcicki_2021,Efimov:21}. On the other hand, the main difficulty encountered by 3D semi-classical studies of multi-electron ionization that include the Coulomb singularity    is unphysical autoionization. Namely, one of the bound electrons can undergo a close encounter with the core and acquire a very negative energy leading   to the escape of another bound electron. This is avoided in quantum mechanical treatments of multi-electron systems due to the lower energy bound of an electron.  Adding a Heisenberg potential is an approach adopted to exclude unphysical autoionization in 3D semi-classical treatments     \cite{PhysRevA.21.834}.  This potential amounts to adding a potential barrier that mimics the Heisenberg uncertainty principle and prevents each electron from a close encounter with the core.   The addition to the Hamiltonian of an extra momentum and position dependent term  results in the momentum of a particle being no longer directly related to the rate of change of its position, $\mathbf{p} \neq m\dot{\mathbf{r}}$ \cite{PhysRevA.51.266,PhysRevA.54.573}. In what follows, we refer to this model as H-model. An advantage of this model is that  electronic interactions   are accounted for with Coulomb forces at all times during propagation. However, due to the Heisenberg potential, each electron accesses  a reduced phase space resulting  in a less accurate description of the interaction of each electron with the core. Indeed, in what follows we show that the H-model   gives rise to ``softer" re-collisions upon the return of an electron to the core.

 
  Here, we take another approach to addressing unphysical autoionization in 3D semi-classical models that include the Coulomb singularity. We develop a 3D semi-classical model that describes the interaction between a pair of bound electrons via an effective Coulomb potential \cite{PhysRevA.40.6223}. The interaction between all other pairs of electrons are described with Coulomb forces. This model advances our previous work of triple ionization  in strongly-driven $\mathrm{H_2He^{+}}$, where we switched off the Coulomb force between bound electrons \cite{PhysRevA.103.043109}.  In the current work, we develop an efficient set of criteria  to determine on the fly, i.e. during time-propagation whether  an electron is bound or ``quasi-free". Hence, we determine on the fly whether the interaction  between two electrons is described by the Coulomb or the effective Coulomb potential.    We refer to this model as ECBB---effective Coulomb potential for bound-bound electrons. We show that  the ECBB-model accurately describes    three- and two-electron ionization spectra  in strongly-driven three-electron atoms. 

Our motivation for developing the ECBB-model is the accurate description at all times of the Coulomb interaction of each electron with the core, unlike the H-model. The importance of this interaction has been demonstrated in the finger-like structure in  the correlated electron momenta in double ionization of strongly-driven Helium. This structure was predicted theoretically \cite{PhysRevLett.96.133001}, observed  experimentally \cite{PhysRevLett.99.263002,PhysRevLett.99.263003} and explained theoretically within a classical framework \cite{PhysRevA.78.023411,PhysRevLett.101.233003}. On the other hand,  the H-model accurately accounts for the interaction between all pairs of electrons with  Coulomb forces.  The ECBB-model does so for pairs of electrons where at least one electron is ``quasi-free". In the ECBB-model the interaction between two bound electrons  is described by effective Coulomb potentials and is thus less accurate. However, bound electrons have a restricted dynamics compared to ``quasi-free" electrons. Hence, one can argue that it is more important to accurately describe the interaction of each electron with the core rather than the interaction between bound electrons.

Here, we formulate the ECBB-model and employ the H-model fully accounting for both the motion of the core and all three electrons and for the magnetic field component of the Lorentz force. That is, we formulate  both models in the non-dipole approximation. This is unlike  previous theoretical studies of strongly-driven atoms. Our formalism is general and can be applied to treat multi-electron ionization in more than three-electron strongly-driven atoms.  Here, we employ both models in the context of strongly-driven Ar. We note that our 3D semi-classical model for two-electron atoms     has previously yielded very good agreement with experimental observables for double ionization in strongly-driven Ar driven by few-cycles laser pulses \cite{ChenA2017Ndiw}. In this latter model of double ionization, we   did not need to address unphysical autoionization. Moreover, we discuss in detail the differences of the ECBB- and H-model  concerning triple and double ionization observables. Such spectra include   the probability distribution of the sum of the electron momenta components along the direction of the laser field  and the correlated electron momenta. Also, we compare our results for the sum of the momenta     with experimental ones \cite{Herrwerth_2008,Zrost_2006}. Finally, we obtain the probability distributions of the angle of escape between two electrons and between an electron and the core. 

 

\section{Method}
In what follows, we describe in detail the formulation of the ECBB-model and the H-model that address multi-electron escape in strongly-driven atoms. The two methods resolve in a  different way  unphysical autoionization in 3D semi-classical models that fully account for  the Coulomb singularity. In both methods,  we propagate in time all three electrons and the core. Moreover, we formulate both methods in the non-dipole approximation fully accounting for the magnetic field component of the laser field.  The Hamiltonian of a four-body system   in the non-dipole approximation is given by
\begin{equation}\label{Ham}
\mathrm{H} = \sum_{\mathrm{i}=1}^{\mathrm{N}}\frac{\left[\mathbf{\tilde{p}}_{\mathrm{i}}- \mathrm{Q_i}\mathbf{A}(\mathbf{r}_{\mathrm{i}},\mathrm{t}) \right]^2}{2\mathrm{m_i}}+\sum_{\mathrm{i}=1}^{\mathrm{N-1}}\sum_{\mathrm{j}=\mathrm{i}+1}^{\mathrm{N}}\frac{\mathrm{Q_i}\mathrm{Q_j}}{|\mathbf{r_i}-\mathbf{r_j}|},
\end{equation}
 where $\mathrm{Q_i}$ is the charge, $\mathrm{m_i}$ is the mass, $\mathbf{r}_{\mathrm{i}}$ is the position vector and $\mathbf{\tilde{p}}_{\mathrm{i}}$ is the canonical momentum vector of particle i. The mechanical momentum $\mathrm{\mathbf{p}_{i}}$ is given by
\begin{equation}
\mathrm{\mathbf{p}_{i}=\mathbf{\tilde{p}}_{\mathrm{i}}- \mathrm{Q_i}\mathbf{A}(\mathbf{r}_{\mathrm{i}},\mathrm{t}}),
\end{equation}
where $\mathrm{\mathbf{A}(\mathbf{r}_{\mathrm{i}},\mathrm{t}})$ is the vector potential and $\mathrm{\mathbf{E}(\mathbf{r}_{\mathrm{i}},\mathrm{t}}) = -\mathrm{\frac{\partial\mathbf{A}(\mathbf{r}_{\mathrm{i}},\mathrm{t})}{\partial t}}$ is the electric field. Modifying  \eq{Ham},  in the following sections, we  formulate the Hamiltonian for   the ECBB- and the H-model. For three-electron Ar, the charge of the core is equal to $\mathrm{Q_1=3}$ a.u. while the mass of the core is equal to $\mathrm{m_1} = 
72820.8 $ a.u.

\subsection{Global regularisation}\label{Regular}
In both methods, we perform a global regularisation to avoid any numerical issues arising from the Coulomb singularities. For strongly-driven H$_2$, we previously used this regularisation scheme to study  double  and ``frustrated" double ionization  within the dipole approximation \cite{toolkit2014} as well as non-dipole effects in non-sequential double ionization \cite{PhysRevA.103.033115}. In this scheme, we define the relative position between two particles i and j as
\begin{equation}
\mathbf{q}_{\mathrm{ij}}=\mathbf{r}_{\mathrm{i}}-\mathbf{r}_{\mathrm{j}}
\label{eq:position}
\end{equation}
and
\begin{equation}
\boldsymbol{\rho}_{\mathrm{ij}}=\frac{1}{\mathrm{N}}\left( \mathbf{\tilde{p}}_{\mathrm{i}} - \mathbf{\tilde{p}}_{\mathrm{j}}- \frac{\mathrm{m_i}-\mathrm{m_j}}{\mathrm{M}}\langle \boldsymbol{\rho} \rangle \right) ,
\end{equation}
where
\begin{equation}
\langle \boldsymbol{\rho} \rangle = \sum_{\mathrm{i=1}}^{\mathrm{N}}\mathbf{\tilde{p}}_{\mathrm{i}} \; \; \text{and} \; \; \mathrm{M}=\sum_{\mathrm{i=1}}^{\mathrm{N}}\mathrm{m_i}.
\end{equation}
The inverse transformation is given by
\begin{equation}\label{position_in_terms_of_q}
\mathbf{r}_\mathrm{i}=\frac{1}{\mathrm{M}}\sum_{\mathrm{j=i+1}}^{\mathrm{N}}\mathrm{m_j}\mathbf{{q}}_{\mathrm{ij}}-\frac{1}{\mathrm{M}}\sum_{\mathrm{j=1}}^{\mathrm{i-1}}\mathrm{m_j}\mathbf{q}_{\mathrm{ji}}+ \langle \mathbf{q} \rangle,
\end{equation}
and
\begin{equation}\label{momenta_in_terms_of_rho}
\mathbf{\tilde{p}}_{\mathrm{i}}=\sum_{\mathrm{j=i+1}}^{\mathrm{N}}\boldsymbol{\rho}_{\mathrm{ij}}-\sum_{\mathrm{j=1}}^{\mathrm{i-1}}\boldsymbol{\rho}_{\mathrm{ji}}+ \frac{\mathrm{m_i}}{\mathrm{M}} \langle \boldsymbol{\rho} \rangle,
\end{equation}
where
\begin{equation}
\langle \mathbf{q} \rangle = \frac{1}{\mathrm{M}} \sum_{\mathrm{i=1}}^{\mathrm{N}}\mathrm{m_i}\mathbf{r}_{\mathrm{i}}.
\end{equation}

Next, we define a fictitious particle $\mathrm{k}$ for each pair of particles $\mathrm{i,j}$ as follows
\begin{equation}
\mathrm{k(i,j)} = \mathrm{ (i-1)N - \dfrac{i(i+1)}{2}+ j },
\label{eqn::kdef}
\end{equation}
with j$>$i and the total number of fictitious particles being equal to $\mathrm{K=N(N-1)/2}$. In addition, we define the parameters $\alpha_{\mathrm{ik}}$ and $\beta_{\mathrm{ik}},$ as $\alpha_{\mathrm{ik}}=1,\beta_{\mathrm{ik}}=\mathrm{m_j/M}$ and $\alpha_{\mathrm{jk}}=-1,\beta_{\mathrm{jk}}=-\mathrm{m_i/M}$ when $\mathrm{k=k(i,j)}$, otherwise $\alpha_{\mathrm{ik}}=\beta_{\mathrm{ik}}=0$. Given the above, Eqs. (\ref{position_in_terms_of_q})  and (\ref{momenta_in_terms_of_rho}) take the following simplified form
\begin{equation}\label{momenta_in_terms_of_rho_v2}
\mathbf{\tilde{p}}_{\mathrm{i}}=\sum_{\mathrm{k=1}}^{K}\alpha_{ik}\boldsymbol{\rho}_{\mathrm{k}}+ \frac{\mathrm{m_i}}{\mathrm{M}} \langle \boldsymbol{\rho} \rangle,
\end{equation}
and
\begin{equation}\label{positions_in_terms_of_rho_v2}
\mathbf{r}_{\mathrm{i}}=\sum_{\mathrm{k=1}}^{K}\beta_{ik}\mathbf{q}_{\mathrm{k}}+ \langle \mathbf{q} \rangle.
\end{equation}

\subsection{Heisenberg potential method}\label{heisenberg_method}
\subsubsection{Description of the model}
The Heisenberg potential, originally proposed by Kirschbaum and Wilets in Ref. \cite{PhysRevA.21.834}, is given by
\begin{equation}\label{Heisenberg}
\mathrm{V_{H,i}}=\dfrac{\xi^2}{4\alpha \mu \mathrm{r}_{i,1}^2}\exp \left\{ \alpha \left[ 1 - \left(  \dfrac{ \mathrm{r}_{i,1} \mathrm{p}_{i,1} }{\xi}  \right)^4 \right]  \right\},
\end{equation}
where $ \mathrm{\mathbf{r}_{i,1}=\mathbf{r}_1 - \mathbf{r}_i } $ is the relative position of each one of the three electrons i=2,3,4 with respect to the core i=1, $\mathrm{\mathbf{p}_{i,1}}$ is the corresponding relative momentum
\begin{equation}
\mathrm{\mathbf{p}_{i,1} = \dfrac{m_i\mathbf{p}_1 - m_1\mathbf{p}_i}{m_1 + m_i}},
\end{equation}
 and $\mu = \mathrm{m_1m_i/(m_i+m_1)} $ is the reduced mass of the electron-core system. 
This potential restricts the relative position and momentum of electron i according to
\begin{equation}\label{Constraint}
\mathrm{r}_\mathrm{i,1}\mathrm{p}_\mathrm{i,1}  \geq \xi.
\end{equation}
Hence, the Heisenberg potential acts as a repulsive potential when the electron is close to the core.

\subsubsection{Hamilton's equations of motion}
Including the Heisenberg potential for each one of the electron-core pairs, the Hamiltonian is given by 
\begin{equation}\label{Hamiltonian_heis}
\mathrm{H} = \sum_{\mathrm{i}=1}^{\mathrm{N}}\frac{\left[\mathbf{\tilde{p}}_{\mathrm{i}}- \mathrm{Q_i}\mathbf{A}(\mathbf{r}_{\mathrm{i}},\mathrm{t}) \right]^2}{2\mathrm{m_i}}+\sum_{\mathrm{i}=1}^{\mathrm{N-1}}\sum_{\mathrm{j}=\mathrm{i}+1}^{\mathrm{N}}\frac{\mathrm{Q_i}\mathrm{Q_j}}{|\mathbf{r}_{\mathrm{i}}-\mathbf{r}_{\mathrm{j}}|} + \sum_{\mathrm{i=2}}^{N}\mathrm{V_{H,i}}.
\end{equation}
Substituting \eq{eq:position} and \eq{momenta_in_terms_of_rho_v2}   in \eq{Hamiltonian_heis}, we obtain the Hamiltonian 
in regularized coordinates as follows
\allowdisplaybreaks{
\begin{align}\label{Hreg}
\mathrm{H} =& \mathrm{\sum_{\mathrm{k,k'=1}}^{\mathrm{K}}\mathrm{T}_{\mathrm{kk'}}\boldsymbol{\rho}_{\mathrm{k}}\boldsymbol{\rho}_{\mathrm{k'}}+\frac{\langle \boldsymbol{\rho}\rangle^2}{2M}+\sum_{\mathrm{k=1}}^{\mathrm{K}}\frac{\mathrm{U_k}}{\mathrm{q_k}}} \nonumber \\
&\mathrm{+\sum_{\mathrm{i=1}}^{\mathrm{N}}\frac{\mathrm{Q^2_i}}{2\mathrm{m_i}}\mathbf{A}^2\left( \mathbf{r}_{\mathrm{i}} ,\mathrm{t} \right)  - \sum_{\mathrm{i=1}}^{\mathrm{N}}\frac{\mathrm{Q_i}}{\mathrm{m_i}}\mathbf{\tilde{p}}_{\mathrm{i}} \cdot \mathbf{A}\left( \mathbf{r}_{\mathrm{i}} ,\mathrm{t} \right)}\\
&\mathrm{+\sum_{i=1}^{N-1}\dfrac{\xi^2}{4\alpha \mu  \mathrm{q}_{i}^2}\exp \left\{ \alpha \left[ 1 - \left(  \dfrac{ \mathrm{q}_{i} p_{i+1,1} }{\xi}  \right)^4 \right]  \right\}} \nonumber, 
\end{align}}
where $\mathbf{\tilde{p}},\mathbf{r}$ are expressed in terms of $\boldsymbol{\rho}$ and $\mathbf{q}$ via  Eqs.  \eqref{momenta_in_terms_of_rho_v2} and  \eqref{positions_in_terms_of_rho_v2}. In \eq{Hreg}, $\mathrm{U_k}$ is equal to $\mathrm{Q_i Q_j}.$
Using \eq{Hreg}, we obtain Hamilton's equations of motion 
\begin{widetext}
\begin{align}\label{eq:new_Equations_of_motion}
\begin{split}
&\frac{\mathrm{d}\mathbf{q}_{\mathrm{k}}}{\mathrm{dt}}=2\sum_{\mathrm{k'}=1}^{\mathrm{K}}\mathrm{T}_{\mathrm{kk'}}\boldsymbol{\rho}_{\mathrm{k'}}-\sum_{\mathrm{i=1}}^{\mathrm{N}}\frac{\mathrm{Q_i}}{\mathrm{m_i}} \alpha_{\mathrm{ik}}\mathbf{A}\left( \mathbf{r}_{\mathrm{i}} ,\mathrm{t} \right) -\mathrm{\sum_{i=1}^{N-1} \dfrac{\left( q_i p_{i+1,1} \right)^2}{\mu \xi^2}}\mathrm{\dfrac{m_{i+1}\alpha_{1k}-m_1\alpha_{i+1k}}{m_1+m_{i+1}}\exp \left\{ \alpha \left[ 1 - \left(  \dfrac{ \mathrm{q}_{i} p_{i+1,1} }{\xi}  \right)^4 \right]  \right\} \mathbf{p}_{i+1,1} } \\
&\frac{\mathrm{d}\langle \mathbf{q} \rangle}{\mathrm{dt}}=\dfrac{1}{\mathrm{M}}\langle \boldsymbol{\rho} \rangle  -\sum_{\mathrm{i=1}}^{\mathrm{N}}\frac{\mathrm{Q_i}}{\mathrm{M}} \mathbf{A}\left( \mathbf{r}_{\mathrm{i}} ,\mathrm{t} \right)\\
&\frac{\mathrm{d}\boldsymbol{\rho}_{\mathrm{k}}}{\mathrm{dt}}= \frac{\mathrm{U_k}\mathbf{q}_{\mathrm{k}}}{\mathrm{q^3_k}} + \sum_{\mathrm{i=1}}^{\mathrm{N}}\frac{\mathrm{Q_i}}{\mathrm{m_i}} \left[ \mathbf{\tilde{p}}_{\mathrm{i}} - \mathrm{Q_i}\mathbf{A}\left( {\mathbf{r}}_{\mathrm{i}},\mathrm{t} \right)\right]\cdot\dfrac{\partial\mathbf{A}\left( {\mathbf{r}}_{\mathrm{i}},\mathrm{t} \right)}{\partial \mathbf{q_k}} \mathrm{+\sum_{i=1}^{N-1}\left[ \dfrac{\xi^2}{2\alpha \mu q_i^4} + \dfrac{\left(p_{i+1,1} \right)^4}{\mu \xi^2} \right] \exp \left\{ \alpha \left[ 1 - \left(  \dfrac{ \mathrm{q}_{i} p_{i+1,1} }{\xi}  \right)^4 \right]  \right\} \mathbf{q_i}\delta_{ik} } \\
 &\mathrm{-\sum_{i=1}^{N-1} \dfrac{\left( q_i p_{i+1,1} \right)^2 }{\mu \left( m_1 + m_{i+1} \right)  \xi^2} \exp \left\{ \alpha \left[ 1 - \left(  \dfrac{ \mathrm{q}_{i} p_{i+1,1} }{\xi}  \right)^4 \right]  \right\} }\mathrm{\mathbf{p}_{i+1,1} \cdot \left[ m_{i+1}Q_1\dfrac{\partial\mathbf{A}\left( {\mathbf{r}}_{1},t \right)}{\partial\mathbf{q}_k } - m_{1}Q_{i+1}\dfrac{\partial\mathbf{A}\left( {\mathbf{r}}_{i+1},t \right)}{\partial\mathbf{q}_k } \right] }\\
&\frac{\mathrm{d}\langle \boldsymbol{\rho}\rangle}{\mathrm{dt}}=\sum_{\mathrm{i=1}}^{\mathrm{N}}\frac{\mathrm{Q_i}}{\mathrm{m_i}} \left[ \mathbf{\tilde{p}}_{\mathrm{i}} - \mathrm{Q_i}\mathbf{A}\left( {\mathbf{r}}_{\mathrm{i}},\mathrm{t} \right)\right]\cdot\dfrac{\partial\mathbf{A}\left( {\mathbf{r}}_{\mathrm{i}},\mathrm{t} \right)}{\partial \langle \mathbf{q} \rangle} \mathrm{-\sum_{i=1}^{N-1} \dfrac{\left( q_i p_{i+1,1} \right)^2 }{\mu \left( m_1 + m_{i+1} \right)  \xi^2} \exp \left\{ \alpha \left[ 1 - \left(  \dfrac{ \mathrm{q}_{i} p_{i+1,1} }{\xi}  \right)^4 \right]  \right\}  } \\
&\mathrm{\times \mathbf{p}_{i+1,1} \cdot \left[ m_{i+1}Q_1\dfrac{\partial\mathbf{A}\left( {\mathbf{r}}_{1},t \right)}{\partial \langle \mathbf{q} \rangle } - m_{1}Q_{i+1}\dfrac{\partial\mathbf{A}\left( {\mathbf{r}}_{i+1},t \right)}{\partial \langle \mathbf{q} \rangle} \right] }.\\
\end{split}
\end{align}
\end{widetext}

\subsubsection{Propagation technique}
To integrate the equations of motion in \eq{eq:new_Equations_of_motion}, we use a leapfrog technique  \citep{Pihajoki2015,Liu2016} jointly with the Bulirsch-Stoer method \cite{press2007numerical,bulirsch1966numerical}. We have previously developed this leapfrog technique to study non-dipole effects in non-sequential double ionization in strongly driven $\mathrm{H_2}$ \cite{PhysRevA.103.033115}.  This leapfrog technique allows to solve Hamilton's equations when the derivative of the position and the momentum depends on the quantities themselves. It is an extension of the leapfrog technique we employed for strongly driven two-electron molecules in the dipole approximation \cite{toolkit2014}. In the latter case the derivative of the position and the momenta do not depend on themselves.

\subsection{Effective Coulomb potential method}\label{effective_method}
\subsubsection{Derivation of the effective Coulomb potential}\label{sec:derivation_eff}
In what follows, we formulate a method that avoids unphysical  autoionization between two bound electrons. To do so, we describe the interaction between two bound electrons with an effective Coulomb potential.
However, we describe the interaction between a ``quasi-free" and a bound electron as well as between two quasi-free electrons with the full Coulomb potential.   In the next subsection, we define the time when an electron transitions from bound to quasi-free and from quasi-free to bound. 

The effective Coulomb potential that electron i experiences due to the charge $\mathrm{\zeta_j}$ of electron j, denoted by $\mathrm{V_{eff}(\zeta_j,|\mathbf{r}_{1}-\mathbf{r}_{i}|)}$, is derived as follows \cite{PhysRevA.40.6223}. We approximate the wavefunction of a bound electron j with a 1s hydrogenic wavefunction
\begin{equation}
\mathrm{\psi(\zeta_j,|\mathbf{r}_{1}-\mathbf{r}_{j}|) = \left( \frac{\zeta_j^3}{\pi} \right)^{1/2} e^{-\zeta_j |\mathbf{r}_{1}-\mathbf{r}_{j}|}},
\end{equation} 
 where the parameter $\mathrm{\zeta_j}$ is later defined in 
Eq. (\ref{eqn::zeta_and_energy}).
The electric charge contained within a sphere of radius $\mathrm{r}$ from the core is given by 
\begin{align}
\begin{split}
\mathrm{Q(\zeta_j,r)}  &= -\mathrm{\int \int \int  \mathrm{|\psi(\zeta_j,r)|^2} dV},
\end{split}
\label{eqn::charge1}
\end{align}
where dV is the volume element in spherical coordinates. 

Using Gauss's law, we find that the effective Coulomb potential that an electron i experiences at a distance $\mathrm{|\mathbf{r}_{1}-\mathbf{r}_{i}|}$ from the core due to the charge distribution of electron j is equal to (see Appendix \ref{appendix:derivation})
\begin{equation}
\mathrm{V_{eff}(\zeta_j,|\mathbf{r}_{1}-\mathbf{r}_{i}|)} =  \mathrm{\dfrac{1-(1+\zeta_j| \mathbf{r}_{1}-\mathbf{r}_{i}|)e^{-2\zeta_j| \mathbf{r}_{1}-\mathbf{r}_{i}|}}{| \mathbf{r}_{1}-\mathbf{r}_{i}|} }. 
\end{equation}
$\mathrm{V_{eff}(\zeta_j,|\mathbf{r}_{1}-\mathbf{r}_{i}|)}$ is a repulsive potential which has limiting values of $\mathrm{\zeta_j}$ when $\mathrm{ |\mathbf{r}_{1}-\mathbf{r}_{i}|=0 }$ and 0 when $\mathrm{ |\mathbf{r}_{1}-\mathbf{r}_{i}|} \rightarrow \infty.$ If the effective charge $\mathrm{\zeta_j(t)}$ is zero then the effective potentials $\mathrm{V_{eff}(\zeta_j,|\mathbf{r}_{1}-\mathbf{r}_{i}|})$ is  zero. The effective charge $\mathrm{\zeta_{j}(t)}$, at any time during the propagation of the four-body system, is proportional to the energy $\mathrm{\mathcal{E}_{j}(t)}$ of electron j, assuming electron j is bound with an energy greater than a lower limit. We set this lower limit to be equal to the ground state energy $\mathcal{E}_{1 s}$ of a hydrogenic atom with core charge equal to  $\mathrm{Q_1}$, i.e. $\mathrm{\mathcal{E}_{1 s} = \frac{Q^2_1}{2}}$. Moreover, when the energy of electron j, $\mathrm{\mathcal{E}_{j}(t)}$, is greater than zero, we set $\mathrm{\zeta_{j}(t)}$ equal to zero, while if the energy is less than the lower limit $\mathcal{E}_{1 s}$ we set $\mathrm{\zeta_{j}(t)}$  equal to $\mathrm{Q_1}$. Hence, we define $\mathrm{\zeta_{j}(t)}$ as follows
\begin{equation}\label{eqn::zeta_and_energy}
\mathrm{\zeta_{j}(t)} = \left\{
    \begin{array}{ll}
        \mathrm{Q_1} & \mathcal{E}_{\mathrm{j}}\mathrm{(t)}  \leq \mathcal{E}_{\mathrm{1 s}}\\
        \mathrm{-\left(\mathrm{Q_1} / \mathcal{E}_{1 s}\right) \mathcal{E}_{j}(t)} & \mathcal{E}_{\mathrm{1 s}} < \mathcal{E}_{\mathrm{j}}\mathrm{(t)}  < 0 \\
        0 & \mathcal{E}_{\mathrm{j}}\mathrm{(t)}  \geq 0,
    \end{array}
\right.
\end{equation}
where the energy $\mathrm{\mathcal{E}_{j}(t)}$ of electron j is given by
\begin{align}\label{eq:energy_of_electron_j}
\mathrm{\mathcal{E}_{j}(t) }&= \mathrm{ \frac{\left[\mathbf{\tilde{p}}_{\mathrm{j}}- \mathrm{Q_j}\mathbf{A}(\mathbf{r}_{\mathrm{j}},\mathrm{t}) \right]^2}{2\mathrm{m_j}} + \frac{Q_j Q_1}{|\mathbf{r}_{1}-\mathbf{r}_{j}|}  - Q_j\mathrm{\mathbf{r}_{j} \cdot \mathbf{E}\left(\mathbf{r}_{j}, t\right)} } \nonumber \\
 &+ \mathrm{\sum_{\substack{\;\mathrm{i=2} \\ \mathrm{i} \neq \mathrm{j}}}^{\mathrm{N}} \mathrm{c_{i,j}(t)}\mathrm{V_{eff}(\zeta_i,|\mathbf{r}_{1}-\mathbf{r}_{j}|})}.
\end{align}
The functions $\mathrm{c_{i,j}(t)}$ determine whether the full Coulomb interaction or the effective $\mathrm{V_{eff}(\zeta_i,|\mathbf{r}_{1}-\mathbf{r}_{j}|})$ and $\mathrm{V_{eff}(\zeta_j,|\mathbf{r}_{1}-\mathbf{r}_{i}|})$ potential interactions are on or off for any pair of electrons i and j during the time propagation. Specifically, the limiting values of $\mathrm{c_{i,j}(t)}$ are zero and one. The value zero corresponds to the full Coulomb potential being turned on while the effective Coulomb potentials are off. This occurs for a pair of electrons i and j where either i or j is quasi-free. The value one corresponds to the effective Coulomb potentials $\mathrm{V_{eff}(\zeta_i,|\mathbf{r}_{1}-\mathbf{r}_{j}|})$ and $\mathrm{V_{eff}(\zeta_j,|\mathbf{r}_{1}-\mathbf{r}_{i}|})$ being turned on while the full Coulomb potential is off. This occurs for bound electrons i and j. For simplicity, we choose $\mathrm{c_{i,j}(t)}$ to change linearly with time between the limiting values zero and one. Hence, $\mathrm{c_{i,j}(t)}$ is defined as follows 
\begin{equation}\label{eqn::zeta_charges_section}
\mathrm{c_{i,j}(t)} = \left\{
    \begin{array}{ll}
        0 & \mathrm{c(t)}  \leq 0 \\
        \mathrm{c(t) } & 0 < \mathrm{c(t)} < 1 \\
        1 & \mathrm{c(t)}  \geq 1,
    \end{array}
\right.
\end{equation}
where $\mathrm{c(t) = \beta (t-t^{i,j}_s)+c_{0},} $ and $\mathrm{c_0}$ is the value of  $\mathrm{c_{i,j}(t)} $ just before a switch at time $\mathrm{t^{i,j}_s}$. A switch at time $\mathrm{t^{i,j}_s}$ occurs if the interaction between electrons i, j changes from full Coulomb to effective Coulomb potential or vice versa. At the start of the propagation at time $\mathrm{t_0},$ $\mathrm{t^{i,j}_s}$ is equal to $\mathrm{t_0}$ and $\mathrm{c_0}$ is one for pairs of electrons that are bound and zero otherwise. To allow for a smooth switch on or switch off of the effective Coulomb potential we choose $\beta$ equal to $\pm 0.1;$ plus corresponds to a switch on and minus to a switch off of the effective Coulomb potential.

\subsubsection{Derivation of the time derivative of the effective charges}\label{sec:derivation_charges_eff}
Including the effective Coulomb potentials, the Hamiltonian of the four-body system is given by
\begin{equation}\label{Hamiltonian_effective}
\begin{aligned}
\mathrm{H} &= \sum_{\mathrm{i}=1}^{\mathrm{N}}\frac{\left[\mathbf{\tilde{p}}_{\mathrm{i}}- \mathrm{Q_i}\mathbf{A}(\mathbf{r}_{\mathrm{i}},\mathrm{t}) \right]^2}{2\mathrm{m_i}}+\sum_{\mathrm{i}=2}^{\mathrm{N}}\frac{\mathrm{Q_i}\mathrm{Q_1}}{|\mathbf{r_1}-\mathbf{r_i}|} \\
&+\sum_{\mathrm{i}=2}^{\mathrm{N-1}}\sum_{\mathrm{j}=\mathrm{i}+1}^{\mathrm{N}} \left[ \mathrm{1-c_{i,j}(t)}\right]\frac{\mathrm{Q_i}\mathrm{Q_j}}{|\mathbf{r_i}-\mathbf{r_j}|} \\
&+\sum_{\mathrm{i}=2}^{\mathrm{N-1}}\sum_{\mathrm{j}=\mathrm{i}+1}^{\mathrm{N}}\mathrm{c_{i,j}(t)} \left[ \mathrm{V_{eff}(\zeta_j,|\mathbf{r}_{1}-\mathbf{r}_{i}|)} + \mathrm{V_{eff}(\zeta_i,|\mathbf{r}_{1}-\mathbf{r}_{j}|)}\right] 
\end{aligned}
\end{equation}
The dipole term $\mathrm{  - Q_j\mathbf{r}_{j} \cdot \mathbf{E}\left(\mathbf{r}_{j}, t\right)} $ of \eq{eq:energy_of_electron_j} involving the electric field does not appear in the Hamiltonian \eqref{Hamiltonian_effective}. There is no contradiction.  Indeed, the gauge-invariant energy of a particle does not always coincide with the gauge-dependent Hamiltonian, as discussed in Ref. \cite{kobe1987,keitel}.
We note that the Hamiltonian in \eq{Hamiltonian_effective} depends not only on positions, momenta and time but also on the effective charges. Since the effective charge $\mathrm{\zeta_j}$ is proportional to the energy  $\mathrm{\mathcal{E}_{j}(t)}$, see \eq{eqn::zeta_and_energy}, it follows that we must obtain the derivative with time of $\mathrm{\mathcal{E}_{j}(t)}$. We note that this is necessary at any time during propagation if at least two electrons are bound. To do so, we apply the chain rule in \eq{eq:energy_of_electron_j} and obtain 
{\allowdisplaybreaks
\begin{widetext}
\begin{align}\label{eq::time_derivative_ej}
\mathrm{\mathcal{\dot{E}}_{j}(t) } &= \mathrm{\frac{\partial \mathcal{E}_{j}(t)}{\partial \mathbf{r}_{\mathrm{j}}}\cdot\dot{\mathbf{r}}_{\mathrm{j}} + \frac{\partial \mathcal{E}_{j}(t)}{\partial \mathbf{\tilde{p}}_{\mathrm{j}}}\cdot \dot{\tilde{\mathbf{p}}}_{\mathrm{j}} + \frac{\partial \mathcal{E}_{j}(t)}{\partial \mathbf{r}_{\mathrm{1}}}\cdot\dot{\mathbf{r}}_{\mathrm{1} } + \mathrm{\sum_{\substack{\;\mathrm{l=2} \\ \mathrm{l} \neq \mathrm{j}}}^{\mathrm{N}}}\frac{\partial \mathcal{E}_{j}(t)}{\partial \zeta_l} \dot{\zeta_l}  }+ \mathrm{\frac{\partial \mathcal{E}_{j}(t)}{\partial t}}
=  \mathrm{\frac{\partial \left[\mathcal{E}_{j}(t) - H\right]}{\partial \mathbf{r}_{\mathrm{j}}}\cdot\dot{\mathbf{r}}_{\mathrm{j}} + \frac{\partial \mathcal{E}_{j}(t)}{\partial \mathbf{r}_{\mathrm{1}}}\cdot\dot{\mathbf{r}}_{\mathrm{1} }  + \mathrm{\sum_{\substack{\;\mathrm{l=2} \\ \mathrm{l} \neq \mathrm{j}}}^{\mathrm{N}}}\frac{\partial \mathcal{E}_{j}(t)}{\partial \zeta_l} \dot{\zeta_l}  }+ \mathrm{\frac{\partial \mathcal{E}_{j}(t)}{\partial t}},\nonumber\\
&=\mathrm{\frac{\partial \left[ - Q_j\mathrm{\mathbf{r}_{j} \cdot \mathbf{E}\left(\mathbf{r}_{j}, t\right)}-\sum_{\mathrm{i}=2}^{\mathrm{N-1}}\sum_{\mathrm{m}=\mathrm{i}+1}^{\mathrm{N}} \left[ \mathrm{1-c_{i,m}(t)}\right]\frac{\mathrm{Q_i}\mathrm{Q_m}}{|\mathbf{r_i}-\mathbf{r_m}|} \right]}{\partial \mathbf{r}_{\mathrm{j}}}\cdot\dot{\mathbf{r}}_{\mathrm{j}}} + \mathrm{\left[-\frac{Q_1 Q_j(\mathbf{r}_{1}-\mathbf{r}_{j})}{|\mathbf{r}_{1}-\mathbf{r}_{j}|^3} + \mathrm{\sum_{\substack{\;\mathrm{i=2} \\ \mathrm{i} \neq \mathrm{j}}}^{\mathrm{N}}} c_{i,j}(t)\frac{\partial V_{eff}(\zeta_i,|\mathbf{r}_{1}-\mathbf{r}_{j}|)}{\partial \mathbf{r}_1} \right] \cdot \dot{\mathbf{r}}_{\mathrm{1}} } \nonumber \\
& +\mathrm{\mathrm{\sum_{\substack{\;\mathrm{i=2} \\ \mathrm{i} \neq \mathrm{j}}}^{\mathrm{N}}}  c_{i,j}(t) \frac{\partial V_{eff}(\zeta_i,|\mathbf{r}_{1}-\mathbf{r}_{j}|)}{\partial \zeta_i} \dot{\zeta_i}  } + \mathrm{\sum_{\substack{\;\mathrm{i=2} \\ \mathrm{i} \neq \mathrm{j}}}^{\mathrm{N}}}\mathrm{\dot{c}_{i,j}(t) V_{eff}(\zeta_i,|\mathbf{r}_{1}-\mathbf{r}_{j}|) + Q_j\mathrm{\mathbf{\dot{r}}_{j} \cdot \mathbf{E}\left(\mathbf{r}_{j}, t\right) - Q_j\mathbf{r}_{j} \cdot \frac{\partial \mathbf{E}\left(\mathbf{r}_{j}, t\right)}{\partial t}}}\\
&= \mathrm{\sum_{\mathrm{i}=2}^{\mathrm{N-1}}\sum_{\mathrm{m}=\mathrm{i}+1}^{\mathrm{N}}[1-c_{i,m}(t)] \mathrm{\frac{Q_i Q_m(\mathbf{r}_{i}-\mathbf{r}_{m})}{|\mathbf{r}_{i}-\mathbf{r}_{m}|^3}\left(\delta_{i,j} - \delta_{m,j}\right)  \cdot \dot{\mathbf{r}}_{\mathrm{j}}  } }   +\mathrm{\left[-\frac{Q_1 Q_j(\mathbf{r}_{1}-\mathbf{r}_{j})}{|\mathbf{r}_{1}-\mathbf{r}_{j}|^3} + \mathrm{\sum_{\substack{\;\mathrm{i=2} \\ \mathrm{i} \neq \mathrm{j}}}^{\mathrm{N}}} c_{i,j}(t)\frac{\partial V_{eff}(\zeta_i,|\mathbf{r}_{1}-\mathbf{r}_{j}|}{\partial \mathbf{r}_1} \right] \cdot \dot{\mathbf{r}}_{\mathrm{1}} } \nonumber\\
+&\mathrm{\mathrm{\sum_{\substack{\;\mathrm{i=2} \\ \mathrm{i} \neq \mathrm{j}}}^{\mathrm{N}}}  \left[c_{i,j}(t) \frac{\partial V_{eff}(\zeta_i,|\mathbf{r}_{1}-\mathbf{r}_{j}|)}{\partial \zeta_i} \dot{\zeta_i} + \dot{c}_{i,j}(t) V_{eff}(\zeta_i,|\mathbf{r}_{1}-\mathbf{r}_{j}|) \right]} -\mathrm{\mathrm{Q_j}\mathbf{r}_j \cdot \dot{\mathbf{E}}(\mathbf{r}_j,\mathrm{t})} \nonumber,
\end{align}
\end{widetext}}
\noindent where we use $\mathrm{\dot{\mathbf{r}}_j = \frac{\partial \mathcal{E}_{j}(t)}{\partial \mathbf{\tilde{p}}_{\mathrm{j}}}}$ and $\dot{\tilde{\mathbf{p}}}_{\mathrm{j}}=-\mathrm{\frac{\partial H }{\partial \mathbf{r}_j }} $. The above expression can be finally written as
\begin{equation}\label{linear}
\mathrm{\mathcal{\dot{E}}_{j}(t) } = \mathrm{f_j} + \mathrm{\mathrm{\sum_{\substack{\;\mathrm{i=2} \\ \mathrm{i} \neq \mathrm{j}}}^{\mathrm{N}}}  c_{i,j}(t) \frac{\partial V_{eff}(\zeta_i,|\mathbf{r}_{1}-\mathbf{r}_{j}|)}{\partial \zeta_i} \dot{\zeta_i} },
\end{equation} 
where $\mathrm{f_j(\mathbf{r},\mathbf{p},t,\mathcal{E})} $ are all the terms in Eq.  (\ref{eq::time_derivative_ej}) that do not depend on $\mathrm{\dot{\zeta}_i(t)}$. The time derivative of $\mathrm{\zeta_i}$ is given by
\begin{equation}\label{zetadot}
\mathrm{\dot{\zeta_i}} = \left\{
    \begin{array}{ll}
        0 & \mathcal{E}_{\mathrm{i}}\mathrm{(t)}  \leq \mathcal{E}_{\mathrm{1 s}}\\
        \mathrm{-\left(\mathrm{Q_1} / \mathcal{E}_{1 s}\right) \dot{\mathcal{E}_{i}}(t)} & \mathcal{E}_{\mathrm{1 s}} < \mathcal{E}_{\mathrm{i}}\mathrm{(t)}  < 0. \\
        0 & \mathcal{E}_{\mathrm{i}}\mathrm{(t)}  \geq 0,
    \end{array}
\right.
\end{equation}

\noindent  We obtain an equation similar to \eq{linear} for each electron. Hence, at any time during propagation, we solve a system of equations to obtain the derivative in time of the energies of each electron. As a result, we express each $\mathrm{\dot{\mathcal{E}}}$ as a function of $\mathrm{(\mathbf{r},\mathbf{p},t,\mathcal{E})}$ with no dependence on the derivatives of the energies.
\subsubsection{Hamilton's equations of motion}

Substituting Eqs. \eqref{eq:position} and \eqref{momenta_in_terms_of_rho_v2} in \eq{Hamiltonian_effective}, we find the Hamiltonian in regularized coordinates to be given by
 \begin{align}\label{Hamiltonian_in_transformed_coordinates}
 \begin{split}
\mathrm{H} =& \mathrm{\sum_{\mathrm{k,k'=1}}^{\mathrm{K}}\mathrm{T}_{\mathrm{kk'}}\boldsymbol{\rho}_{\mathrm{k}}\boldsymbol{\rho}_{\mathrm{k'}}+\frac{\langle \boldsymbol{\rho}\rangle^2}{2M}} + \sum_{\mathrm{k}=1}^{\mathrm{K}}[1-\mathrm{c_{k}(t)}]\frac{\mathrm{U_k}}{\mathrm{q_k}}\\
&\mathrm{+\sum_{\mathrm{i=1}}^{\mathrm{N}}\frac{\mathrm{Q^2_i}}{2\mathrm{m_i}}\mathbf{A}^2\left( \mathbf{r}_{\mathrm{i}} ,\mathrm{t} \right)  - \sum_{\mathrm{i=1}}^{\mathrm{N}}\frac{\mathrm{Q_i}}{\mathrm{m_i}}\mathbf{\tilde{p}}_{\mathrm{i}} \cdot \mathbf{A}\left( \mathbf{r}_{\mathrm{i}} ,\mathrm{t} \right)}\\
&+\sum_{\mathrm{k=1}}^{\mathrm{K}}\mathrm{c_{k}(t)}\mathrm{V_{k}}, 
\end{split}
\end{align}
where
\begin{align}\label{veff_regular}
\begin{split}
\mathrm{V_{k(i,j)}} &= \mathrm{V_{eff}(\zeta_j,|\mathbf{r}_{1}-\mathbf{r}_{i}|)} + \mathrm{V_{eff}(\zeta_i,|\mathbf{r}_{1}-\mathbf{r}_{j}|)},
\end{split}
\end{align}
and $\mathbf{\tilde{p}}$, $\mathbf{r},$ are expressed in terms of $\boldsymbol{\rho}$ and $\mathbf{q}$ via  Eqs. \eqref{momenta_in_terms_of_rho_v2} and \eqref{positions_in_terms_of_rho_v2}. Moreover, for k=1,2,3
 $\mathbf{q}_{\mathrm{k}}$ corresponds to the relative distance between each one of the three electrons and the core. Since, the Coulomb force between each of the three electrons and the core is always on, we set $\mathrm{c_k(t)=0}$, for k=1,2,3. Using \eq{Hamiltonian_in_transformed_coordinates}, we find that Hamilton's equations of motion are given by 
\begin{align}\label{eq:new_Equations_of_motion_eff}
\begin{split}
\frac{\mathrm{d}\mathbf{q}_{\mathrm{k}}}{\mathrm{dt}}&=2\sum_{\mathrm{k'}=1}^{\mathrm{K}}\mathrm{T}_{\mathrm{kk'}}\boldsymbol{\rho}_{\mathrm{k'}}-\sum_{\mathrm{i=1}}^{\mathrm{N}}\frac{\mathrm{Q_i}}{\mathrm{m_i}} \alpha_{\mathrm{ik}}\mathbf{A}\left( \mathbf{r}_{\mathrm{i}} ,\mathrm{t} \right)   \\
\frac{\mathrm{d}\langle \mathbf{q} \rangle}{\mathrm{dt}}&=\dfrac{1}{\mathrm{M}}\langle \boldsymbol{\rho} \rangle -\sum_{\mathrm{i=1}}^{\mathrm{N}}\frac{\mathrm{Q_i}}{\mathrm{M}} \mathbf{A}\left( \mathbf{r}_{\mathrm{i}} ,\mathrm{t} \right)  \\
\frac{\mathrm{d}\boldsymbol{\rho}_{\mathrm{k}}}{\mathrm{dt}}&= \mathrm{[1-c_k(t)] }\frac{\mathrm{U_k}\mathbf{q}_{\mathrm{k}}}{\mathrm{q^3_k}} -\sum_{\mathrm{k'=1}}^{\mathrm{K}}\mathrm{c_{k'}(t)}\mathrm{\frac{\partial V_{k'}}{\partial \mathbf{q_k}}}\\
&+ \sum_{\mathrm{i=1}}^{\mathrm{N}}\frac{\mathrm{Q_i}}{\mathrm{m_i}} \left[ \mathbf{\tilde{p}}_{\mathrm{i}} - \mathrm{Q_i}\mathbf{A}\left( {\mathbf{r}}_{\mathrm{i}},\mathrm{t} \right)\right]\cdot\dfrac{\partial\mathbf{A}\left( {\mathbf{r}}_{\mathrm{i}},\mathrm{t} \right)}{\partial \mathbf{q_k}}  \\
\frac{\mathrm{d}\langle \boldsymbol{\rho}\rangle}{\mathrm{dt}}&=\sum_{\mathrm{i=1}}^{\mathrm{N}}\frac{\mathrm{Q_i}}{\mathrm{m_i}} \left[ \mathbf{\tilde{p}}_{\mathrm{i}} - \mathrm{Q_i}\mathbf{A}\left( {\mathbf{r}}_{\mathrm{i}},\mathrm{t} \right)\right]\cdot\dfrac{\partial\mathbf{A}\left( {\mathbf{r}}_{\mathrm{i}},\mathrm{t} \right)}{\partial \langle \mathbf{q} \rangle},
\end{split}
\end{align}
where
\begin{equation}\label{eq:new_Equations_of_motion_eff_more}
\mathrm{\frac{\partial V_{k'(i,j)}}{\partial \mathbf{q_k}}} =  \mathrm{\frac{\partial V_{k'}}{\partial \mathbf{q_k} }\delta_{k,k(1,i)} + \frac{\partial V_{k'}}{\partial \mathbf{q_k} }\delta_{k,k(1,j)} },
 \end{equation}
 where $\mathbf{\tilde{p}},\mathbf{r}$ are expressed in terms of $\boldsymbol{\rho}$ and $\mathbf{q}$ via  Eqs.  \eqref{momenta_in_terms_of_rho_v2} and \eqref{positions_in_terms_of_rho_v2}. From Eqs. \eqref{eq:new_Equations_of_motion_eff} and \eqref{eq:new_Equations_of_motion_eff_more} it follows that the term $\mathrm{\sum_{k'=1}^{K}c_{k'}(t)\dfrac{\partial V_{k'}}{\partial \mathbf{q_k}}}$ is non zero for k = 1,2,3 and has the following form
\begin{widetext}
\begin{align*}
\mathrm{\sum_{k'=1}^{K}c_{k'}(t)\dfrac{\partial V_{k'}}{\partial \mathbf{q_1}}}&= \mathrm{c_4(t)\dfrac{\partial V_{eff}(\zeta_3,| \mathbf{r}_1 - \mathbf{r}_2 |)}{\partial \mathbf{q_1}}} + \mathrm{c_5(t)\dfrac{\partial V_{eff}(\zeta_4,| \mathbf{r}_1 - \mathbf{r}_2 |)}{\partial \mathbf{q_1}}}\\&=\mathrm{c_4(t)\dfrac{-1+\left[ 1+2\zeta_3q_{1}(1+\zeta_3 q_1)\right] e^{-2\zeta_3\mathrm{q_1}}}{q^3_1}}\mathbf{q_1} + \mathrm{c_5(t)\dfrac{-1+\left[ 1+2\zeta_4q_{1}(1+\zeta_4 q_1)\right] e^{-2\zeta_4\mathrm{q_1}}}{q^3_1}}\mathbf{q_1} \\
\mathrm{\sum_{k'=1}^{K}c_{k'}(t)\dfrac{\partial V_{k'}}{\partial \mathbf{q_2}}}&= \mathrm{c_4(t)\dfrac{\partial V_{eff}(\zeta_2,| \mathbf{r}_1 - \mathbf{r}_3 |)}{\partial \mathbf{q_2}}} + \mathrm{c_6(t)\dfrac{\partial V_{eff}(\zeta_4,| \mathbf{r}_1 - \mathbf{r}_3 |)}{\partial \mathbf{q_2}}}\\&=\mathrm{c_4(t)\dfrac{-1+\left[ 1+2\zeta_2q_{2}(1+\zeta_2 q_2)\right] e^{-2\zeta_2\mathrm{q_2}}}{q^3_2}}\mathbf{q_2} + \mathrm{c_6(t)\dfrac{-1+\left[ 1+2\zeta_4q_{2}(1+\zeta_4 q_2)\right] e^{-2\zeta_4\mathrm{q_2}}}{q^3_2}}\mathbf{q_2} \\
\mathrm{\sum_{k'=1}^{K}c_{k'}(t)\dfrac{\partial V_{k'}}{\partial \mathbf{q_3}}}&= \mathrm{c_5(t)\dfrac{\partial V_{eff}(\zeta_2,| \mathbf{r}_1 - \mathbf{r}_4 |)}{\partial \mathbf{q_3}}} + \mathrm{c_6(t)\dfrac{\partial V_{eff}(\zeta_3,| \mathbf{r}_1 - \mathbf{r}_4 |)}{\partial \mathbf{q_3}}}\\&=\mathrm{c_5(t)\dfrac{-1+\left[ 1+2\zeta_2q_{3}(1+\zeta_2 q_3)\right] e^{-2\zeta_2\mathrm{q_3}}}{q^3_3}}\mathbf{q_3} + \mathrm{c_6(t)\dfrac{-1+\left[ 1+2\zeta_3q_{3}(1+\zeta_3 q_3)\right] e^{-2\zeta_3\mathrm{q_3}}}{q^3_3}}\mathbf{q_3}.
\end{align*}
\end{widetext}
In addition to \eq{eq:new_Equations_of_motion_eff}, we have three more equations for $\mathrm{\dot{\mathcal{E}}(\mathbf{q},\boldsymbol{\rho},t,\mathcal{E})}.$

 \subsubsection{Propagation technique}
 In our formulation, we fully account for the Coulomb singularities. Hence, an electron can approach infinitely close to the nucleus during time propagation. To ensure the accurate numerical treatment of the N-body problem in the laser field, we perform a global regularisation.  This regularisation was introduced in the context of the gravitational N-body problem  \cite{Heggie1974}. Here, we integrate the equations of motion using  a leapfrog technique  \citep{Pihajoki2015,Liu2016} jointly with  the Bulirsch-Stoer method \cite{press2007numerical,bulirsch1966numerical}.  This leapfrog technique allows integration of Hamilton's equation when the  derivatives of the positions and the momenta  depend  on the quantities themselves. We previously employed this technique in our studies of non-dipole effects in non-sequential double ionization of strongly driven $\mathrm{H_2}$ \cite{PhysRevA.103.033115}. The difference between the leapfrog technique employed in this work and the one previously employed in \cite{PhysRevA.103.033115} is that the former is more involved. Indeed, in the current leapfrog technique  we also need to propagate in time the energies $\mathcal{E}(\mathrm{t})$, see  \eq{eq::time_derivative_ej}. The steps involved in this leapfrog technique are as follows.
  
\begin{figure}[t]
\includegraphics[width=\linewidth]{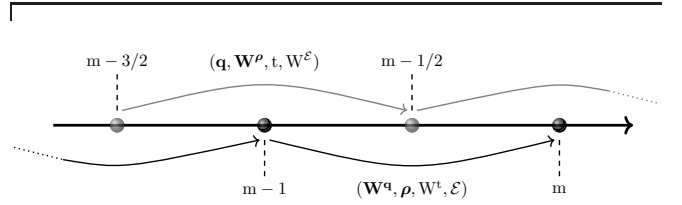}
\caption{Schematic illustration of the propagation of the two quadruplets $(\mathbf{q},\mathbf{W}^{\boldsymbol{\rho}},\mathrm{t},\mathrm{W}^{\mathcal{E}})$ and $(\mathbf{W^q},\boldsymbol{\rho},\mathrm{W^t},\mathcal{E})$ over a sub step of size h, $\mathrm{m-3/2}\to\mathrm{m-1/2}$ and $\mathrm{m-1}\to\mathrm{m}$ respectively, with m=2,...,n-1. }\label{fig:leapfrog_illustration}
\end{figure}

 First,  we perform a time transformation $\mathrm{t}\to\mathrm{s}$, where
\begin{equation}
\mathrm{ds}=\Omega(\mathbf{q})\mathrm{dt},
\end{equation}
with $\Omega(\mathbf{q})$ an arbitrary positive function of $\mathbf{q}.$ We select the function
\begin{equation}
\Omega(\mathbf{q}) = \sum_{\mathrm{k=1}}^{\mathrm{K}}\dfrac{1}{|\mathbf{q}_\mathrm{k}|},
\end{equation}
which forces  the time step to decrease when two particles undergo a close encounter and to increase when all particles are far away from each other. The equations of motion now take the following form
\begin{align}
\begin{split}
\mathbf{q'}&=\dot{\mathbf{q}}(\mathbf{q},\boldsymbol{\rho},\mathrm{t})/\Omega(\mathbf{q})\\
\boldsymbol{\rho'}&=\dot{\boldsymbol{\rho}}(\mathbf{q},\boldsymbol{\rho},\mathrm{t},\mathcal{E})/\Omega(\mathbf{q})\\
\mathrm{t'}&=1/\Omega(\mathbf{q}) \\
\mathcal{E}'&=\dot{\mathcal{E}}(\mathbf{q},\boldsymbol{\rho},\mathrm{t},\mathcal{E})/\Omega(\mathbf{q}),
\end{split}
\end{align}
with prime denoting the derivative with respect to the new variable s. The integration is based on the leapfrog technique described in \citep{PhysRevA.103.033115} that introduces four auxiliary variables, two vectors $\mathbf{W^q},\mathbf{W}^{\boldsymbol{\rho}}$ and two scalars $\mathrm{W^t},\mathrm{W}^{\mathcal{E}}$. 
As a result, an extended system is obtained where the derivatives of the position, the momenta and the energies no longer depend on the quantities themselves.
 The extended equations are given by
\begin{align*}
\begin{split}
\mathbf{q'}&=\dot{\mathbf{q}}(\mathbf{W^q},\boldsymbol{\rho},\mathrm{W^t})/\Omega(\mathbf{W^q})\\
{\mathbf{W}^{\boldsymbol{\rho}}}'&=\dot{\boldsymbol{\rho}}(\mathbf{W^q},\boldsymbol{\rho},\mathrm{W^t},\mathcal{E})/\Omega(\mathbf{W^q})\\
\mathrm{t'}&=1/\Omega(\mathbf{W^q})\\
\mathrm{W^{\mathcal{E}}}'&= \dot{\mathcal{E}}(\mathbf{W^q},\boldsymbol{\rho},\mathrm{W^t},\mathcal{E})/\Omega(\mathbf{W^q}), 
\end{split}
\end{align*}
and
\begin{align*}
\begin{split}
\mathbf{W^{q'}}&=\dot{\mathbf{q}}(\mathbf{q},\mathbf{W^{\boldsymbol{\rho}}},\mathrm{t})/\Omega(\mathbf{q})\\
\boldsymbol{\rho'}&=\dot{\boldsymbol{\rho}}(\mathbf{q},\mathbf{W^{\boldsymbol{\rho}}},\mathrm{t},\mathrm{W}^{\mathcal{E}})/\Omega(\mathbf{q})\\
\mathrm{W^{t'}}&=1/\Omega(\mathbf{q})\\
\mathcal{E}'&= \dot{\mathcal{E}}(\mathbf{q},\mathbf{W}^{\boldsymbol{\rho}},\mathrm{t},\mathrm{W}^\mathcal{E})/\Omega(\mathbf{q}).
\end{split}
\end{align*}
We propagate for a  time step, by propagating for half a step each quadruplet of variables ($\mathbf{q},\mathbf{W}^{\boldsymbol{\rho}}$,t,W$^\mathcal{E}$) and ($\mathbf{W^q},\boldsymbol{\rho},\mathrm{W^t}$,$\mathcal{E}$)  in an alternating way, see the leapfrog algorithm described in   Appendix \ref{AppendixLeapfrog}. 
Moreover, to achieve better accuracy, we incorporate the leapfrog method in the Bulirsch-Stoer extrapolation scheme  \cite{press2007numerical,bulirsch1966numerical}.   In this scheme, a propagation over a step H, is split into n sub steps of size $\mathrm{h=H/n}.$ We use the leapfrog method to propagate over each sub step. In \fig{fig:leapfrog_illustration}, we offer a schematic illustration of the propagation during a time sub step of size h. The detailed algorithm is described in the Appendix \ref{AppendixLeapfrog}.  This process  is repeated with increasing number of sub steps, i.e. n$\to \infty,$ until an extrapolation with a satisfactory error is achieved. 

 
\subsubsection{Definition of quasi-free  and bound electron}\label{switching}
In the ECBB-model the interaction  between a pair of electrons where at least one is quasi-free is described with Coulomb forces. The interaction between bound electrons is  described with effective Coulomb potentials. 
Hence, we need to define during time propagation, i.e. on the fly, if an electron is quasi-free or bound. At the start of propagation, the electron that tunnel-ionises  (electron 2) is considered quasi-free and the other two (electrons 3 and 4) are bound. We denote the core as particle 1.

\begin{figure}[H]
\centering
\includegraphics[width=\linewidth]{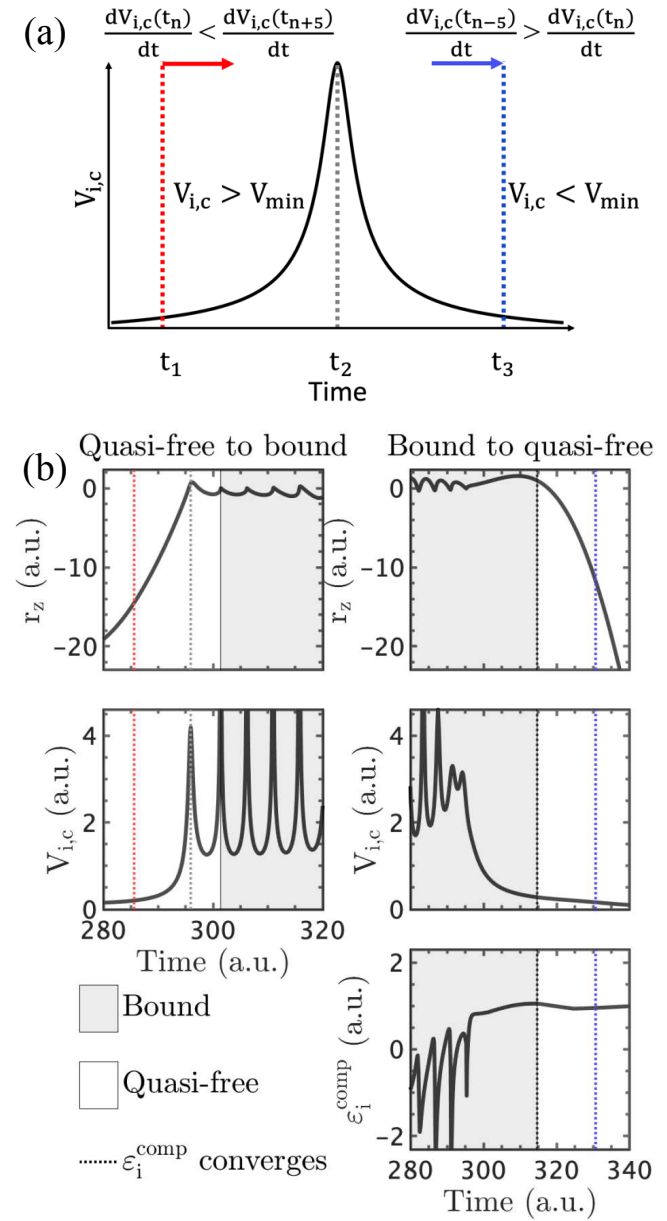}
\caption{Schematic illustration of the criteria to determine when a quasi-free electron becomes bound (left column) and when a bound electron becomes quasi-free (right column).}
\label{fig:recollision}
\end{figure}

At  times $\mathrm{t>t_0}$, a quasi-free electron i transitions to bound if the following conditions are satisfied: (i) the potential of electron i with the core, $\mathrm{V_{i,c}}$,  is larger than a threshold value, i.e. $\mathrm{V_{i,c}> V_{min}}$ at $\mathrm{t_1}$, and $\mathrm{V_{i,c}}$ is continuously increasing, i.e. $\mathrm{\frac{dV_{i,c}(t_{n+5})}{dt} > \frac{dV_{i,c}(t_{n})}{dt} }$ for five times $\mathrm{t_n}$ which are five time steps apart with the first one being at time $\mathrm{t_1}$, see \fig{fig:recollision}(a); (ii) the position of electron i along the electric field, i.e. z axis here,  has at least two extrema of the same kind, i.e. two maxima or two minima, in a time interval less than half a period of the laser field. We start checking if condition (ii) is satisfied at  time $\mathrm{t_2}$ when  electron i has the closest approach to the core, i.e. $\mathrm{V_{i,c}}$ is maximum. We stop checking whether condition (ii) is satisfied at  time $\mathrm{t_3}$ when $\mathrm{V_{i,c}}$ is smaller than the threshold value $\mathrm{V_{min}}$ and $\mathrm{V_{i,c}}$ is continuously decreasing, i.e. $\mathrm{\frac{dV_{i,c}(t_n)}{dt} < \frac{dV_{i,c}(t_{n-5})}{dt} }$ for five times $\mathrm{t_n}$ which are five time steps apart with the last one being at time $\mathrm{t_3}$, see \fig{fig:recollision}(a).   In the current study, we set $\mathrm{V_{min}}$ equal to 3/15 which is equal to 0.2 a.u. We find that our results remain almost the same for a range of values of $\mathrm{V_{min}}$. 
Also, at the end of the laser pulse, we check whether a quasi-free electron has positive or negative compensated energy \cite{Leopold_1979}. If the latter occurs, we consider the electron to be bound.
 Accounting for the effective Coulomb potential, the compensated energy of electron i   is given by
\begin{align}\label{eqn::compensated1}
\varepsilon^{\mathrm{comp}}_{\mathrm{i}}(\mathrm{t})= \mathrm{ \frac{\mathbf{\tilde{p}}^2_{\mathrm{i}} }{2\mathrm{m_i}} + \frac{Q_1 Q_i}{|\mathbf{r}_1-\mathbf{r}_i|}  + \mathrm{\sum_{\substack{\;\mathrm{j=2} \\ \mathrm{j} \neq \mathrm{i}}}^{\mathrm{N}} \mathrm{c_{i,j}(t)}\mathrm{V_{eff}(\zeta_j,|\mathbf{r}_{1}-\mathbf{r}_{i}|})}}.
\end{align}

A bound electron transitions to quasi-free  at time $\mathrm{t>t_0}$ if either one of the following two conditions is satisfied: (i) at time t the compensated energy  of electron i converges to a positive value; (ii) at times $\mathrm{t=t_3}$, $\mathrm{V_{i,c}}$ is smaller than the threshold value $\mathrm{V_{min}}$ and $\mathrm{V_{i,c}}$ is continuously decreasing, i.e. $\mathrm{\frac{dV_{i,c}(t_n)}{dt} < \frac{dV_{i,c}(t_{n-5})}{dt} }$ for five $\mathrm{t_n}$ which are five time steps apart,  the last one being at  $\mathrm{t_3}$.

We illustrate the above criteria in \fig{fig:recollision}(b). We  denote the times $\mathrm{t_1}$, $\mathrm{t_2}$ and $\mathrm{t_3}$ with red, grey and blue vertical dashed lines, respectively. In the left column, we plot the position $\mathrm{r_z}$ and the potential $\mathrm{V_{i,c}}$ of a quasi-free electron as it transitions to bound. In the right column, we plot the position $\mathrm{r_z}$, the potential $\mathrm{V_{i,c}}$ and the compensated energy of a bound electron as it transitions to quasi-free. The black dashed line denotes the time when the compensated energy converges and the electron transitions from bound to quasi-free. For this specific trajectory, the compensated energy converges prior to $\mathrm{t_3}$ and hence electron i transitions from bound to quasi-free at  $\mathrm{t<t_{3}}$.
We note that the criteria for the convergence of the compensated energy and the number of extrema in the position of the electron along the laser field have been used to determine  whether an electron is quasi-free or bound in our previous work on strongly-driven three electron triatomic molecules \cite{PhysRevA.103.043109}. However, the criteria presented above are considerably refined compared to the ones in Ref. \cite{PhysRevA.103.043109}, allowing for the full Coulomb forces to be turned on for a longer time interval. Moreover, in the ECBB-model we account for the interaction between  bound electrons with effective Coulomb potentials, while in Ref. \cite{PhysRevA.103.043109} this interaction was  set equal to zero.  

\subsection{Initial conditions}
\subsubsection{Tunnel-ionizing electron}
In both methods, electron 2 tunnel-ionizes at time $\mathrm{t_0}$ through the field-lowered Coulomb-barrier with a rate that is described by the quantum mechanical Ammosov-Delone-Krainov (ADK) formula \cite{Landau,Delone:91}. To obtain the ADK rate, we use the value of the energy needed to ionize one electron from Ar, i.e. we use Ip$_{1}$ = 0.579 a.u. We find t$_{0}$,  using importance sampling \cite{ROTA1986123} in the time interval [-2$\tau$,2$\tau$] where the electric field is non-zero; $\tau$ is the full width at half maximum of the pulse duration in intensity. The importance sampling distribution is given by the ADK ionization rate. The exit point of  electron 2 is along the direction of the laser field and is computed using parabolic coordinates \cite{HUP1997533}.  The momentum of electron 2 is taken to be equal to zero along the laser field. The transverse momentum is given by a Gaussian distribution which represents the Gaussian-shaped filter with an intensity-dependent width arising from standard tunneling theory \cite{Delone:91,Delone_1998,PhysRevLett.112.213001}. 

\subsubsection{Position and momentum distributions of the bound electrons in the H-model}
In the Heisenberg potential, see \eq{Heisenberg}, for a given $\alpha$, we find the value of $\xi$ that ensures that the minimum of the one-electron Hamiltonian 
\begin{equation}\label{eq:one-e_Hamiltonian}
\mathrm{H_i} = \mathrm{\dfrac{\mathbf{p}_i^2}{2m_i} + \frac{Q_1 Q_i}{ | \mathbf{r}_1 - \mathbf{r}_i  |}+ \dfrac{\xi^2}{4\alpha \mu \mathrm{r}_{i,1}^2}\exp \left\{ \alpha \left[ 1 - \left(  \dfrac{ \mathrm{r}_{i,1} \mathrm{p}_{i,1} }{\xi}  \right)^4 \right]  \right\}}
\end{equation} 
corresponds to the third ionization potential of Ar ($\mathrm{Ip_{3}=1.497}$ a.u.) \cite{PhysRevLett.109.053004,PhysRevA.86.043427,Tong:15}. To minimize \eq{eq:one-e_Hamiltonian} with respect to the relative distance $\mathrm{r_{i,1}}$, we start from the lower limit of the constraint 
\begin{equation}\label{constraint2}
\mathrm{r}_\mathrm{i,1}\mathrm{p}_\mathrm{i,1}  = \xi \Rightarrow \mathrm{p}_\mathrm{i,1} = \xi / \mathrm{r}_\mathrm{i,1} 
\end{equation}
Since the mass of the core $\mathrm{m_1 \gg m_i}$ it follows that $\mathrm{p_{i,1} \approx p_i}$. Hence, \eq{constraint2} can be written as $ \mathrm{p_i= \xi / r_{i,1}} $ and substituting in \eq{eq:one-e_Hamiltonian} we obtain
\begin{equation}\label{eq:one-e_Hamiltonian_2}
\mathrm{H_i = \dfrac{\mathbf{\xi}^2}{2m_ir_{i,1}^2} + \frac{Q_1 Q_i}{\mathrm{r_{i,1}}} + \dfrac{\xi^2}{4\alpha \mu \mathrm{r}_{i,1}^2}}.
\end{equation} 
The minimum of \eq{eq:one-e_Hamiltonian_2} with respect to $\mathrm{r_{i,1}}$, occurs at
\begin{equation}
\mathrm{r_{i,1}^{min}=-\dfrac{ 2 \alpha \mu + m_i }{2\alpha \mu m_i  Q_1 Q_i} \xi^2},
\end{equation}
and the energy is given by
\begin{equation}
\mathrm{H_{i}^{min}=-\dfrac{ \alpha \mu  m_i \left( Q_1 Q_i \right)^2}{ (m_1 + 2 \alpha \mu)\xi^2 }}.
\end{equation}
Setting this energy equal to  $\mathrm{Ip_{3}}$, we find $\xi=1.55$ a.u for $\alpha=2$  and $\xi = 1.63$ a.u. for $\alpha=4$.  Hence, for $\alpha=2$ the electrons access a larger phase space during the time propagation.

\begin{figure}[H]
\centering
\includegraphics[width=\linewidth]{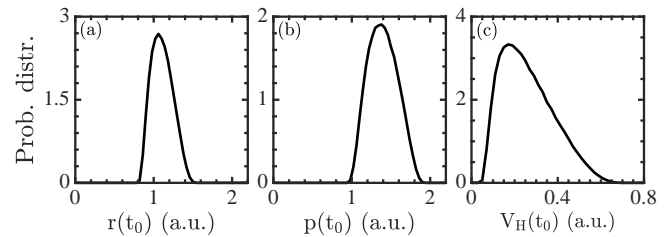}
\caption{ Probability distribution of r (a) and p (b) for each of the electrons 3, 4 as well as the Heisenberg potential $\mathrm{V_{H}}$ (c) at time $\mathrm{t_0}$, for $\alpha=2$.}\label{prob_distribution_t0_heis}
\end{figure}

To find the initial position and momentum vectors of the two initially bound electrons at time $\mathrm{t_{0}}$, we apply a trial and error method similar to the one proposed by Cohen \cite{PhysRevA.54.573}. First, we randomly sample the magnitude of the position and the momentum vector for each electron in the intervals $\mathrm{[0,r_{max}], [0,p_{max}]}$. We find that it is sufficient to consider $\mathrm{r_{max}} = 3$ a.u. and $\mathrm{p_{max}} = 3$ a.u. The $\theta, \phi$ polar and azimuthal angles of the position and the momentum of electrons 3 and 4 are obtained as uniform random numbers of $\cos\theta$ in the interval [-1,1] and $\phi$ in the interval [0,$2\pi$]. Using the position and momenta of electrons 3 and 4, we determine the total energy of the two electrons in the absence of the electric field
\begin{equation}
\mathrm{H_{3,4}} = \mathrm{\sum_{i=3}^{4}\dfrac{\mathbf{p}_i^2}{2} + \sum_{i=3}^{4}\frac{Q_1 Q_i}{\left| \mathbf{r}_1 - \mathbf{r}_i \right|} + \sum_{i=3}^{4}V_{H,i} + \dfrac{Q_3Q_4}{| \mathbf{r}_3 - \mathbf{r}_4  |} }.
\end{equation}
If the energy $\mathrm{H_{3,4}}$ is within 1\% of the binding energy $\mathrm{Ip_{2}+Ip_{3}},$ we accept the initial conditions of electron 3 and 4, otherwise we reject them. For Ar, the energy to ionize a second electron is $\mathrm{Ip_{2}}= 1.015$ a.u. Using the above procedure, we plot in Fig. \eqref{prob_distribution_t0_heis}, the probability distribution of the initial position and momentum of electrons 3 and 4 as well as of the Heisenberg potential.

\subsubsection{Position and momentum distributions of the bound electrons in the ECBB-model}
\begin{figure}[H]
\centering
\includegraphics[width=\linewidth]{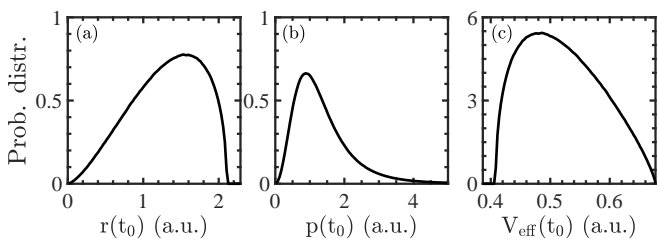}
\caption{Probability distribution of r (a) and p (b) for each of the electrons 3, 4 as well as the effective Coulomb potential $\mathrm{V_{eff}}$ (c) at time $\mathrm{t_0}.$}\label{prob_distribution_t0_eff}
\end{figure}

In the ECBB-model, we obtain the initial position and momentum of electron 4 at time $\mathrm{t_{0}}$ using a microcanonical distribution with an energy 
\begin{align}
\mathrm{\mathcal{E}_{4}(t_0) }= \mathrm{ \frac{\mathbf{p}^2_{\mathrm{4}} }{2\mathrm{m_4}} + \frac{Q_1 Q_4}{|\mathbf{r}_1-\mathbf{r}_4|}  
 + \mathrm{V_{eff}}\left(\zeta_3, | \mathbf{r}_1 - \mathbf{r}_4 |\right)},
\end{align}
and similarly for electron 3. We take the energy $\mathrm{\mathcal{E}_{3}(t_0) = \mathcal{E}_{4}(t_0) = -\mathrm{Ip_{2}}}$ and using \eq{eqn::zeta_and_energy} we find that $\mathrm{\zeta_{3}(t_0)=\zeta_{4}(t_0)}$ = -$\mathrm{\left(Q_{1}/\mathcal{E}_{1 s}\right) \mathrm{\mathcal{E}_{3}(t_0)}}$. The reason we set the initial energy of each electron equal to $-\mathrm{Ip_{2}}$ is that $ \mathcal{E}_{4}(t_0)$ and  $\mathcal{E}_{3}(t_0)$   include the interaction with the other electron via $\mathrm{V_{eff}}$. Hence, $\mathrm{\mathcal{E}_{3}(t_0)}$ and $\mathrm{\mathcal{E}_{4}(t_0)}$ correspond to the energy needed to remove an electron from Ar$^{+}$. Using the above defined microcanonical distribution,  we obtain the initial position and momentum of each bound electron \cite{PhysRevA.33.3859}. In Fig. \eqref{prob_distribution_t0_eff}, we plot the probability distribution for the initial position and momentum of electrons 3 and 4 as well as of the $\mathrm{V_{eff}}$.

\section{Results}
In what follows, we compare observables  for triple ionization (TI)  and double ionization (DI)  obtained with the ECBB-model and the H-model. If available, we also compare these observables with experimental results \cite{Herrwerth_2008,Zrost_2006}.
In our formulation both the ECBB-model and the H-model fully account for non-dipole effects and treat the motion of the electrons and the core  on an equal footing. 

Here, we employ a vector potential of the form
\begin{equation}\label{eq:vector_potential}
\mathbf{A}(\mathrm{y,t}) = -\frac{\mathrm{E}_0}{\omega}\exp \left[ - 2\ln (2)\left( \frac{\mathrm{c t - y}}{\mathrm{c} \tau} \right)^2 \right]   \sin ( \omega \mathrm{t}  - \mathrm{k y}) \hat{\mathbf{z}},
\end{equation}
where $\mathrm{k=\omega/c}$ is the  wave number of the laser field and  $\tau$ is the full width at half maximum of the pulse duration in intensity.  The direction of both the vector potential and the electric field is along the z axis. We take the propagation direction of the laser field to be along the y axis and hence the magnetic field points  along the x axis. We study Ar driven by a laser pulse with intensities ranging  from $\mathrm{2 \times 10^{14} W/cm^2 }$ to $\mathrm{5 \times 10^{14} W/cm^2 }$ and  durations of $\tau = 20$ fs, 25 fs and 30 fs at 800 nm. 



 The time propagation of strongly-driven Ar starts at time $\mathrm{t_0}$ and stops at an asymptotically large time $\mathrm{t_f}$. For each trajectory, if the energies of three (two) electrons are positive, we label the trajectory as a triple (double)  ionization event. The DI and TI probabilities are 
\begin{equation}
\mathrm{P_{DI} = \dfrac{N_{DI}}{N}}, \hspace{0.1cm} \mathrm{P_{TI} = \dfrac{N_{TI}}{N}},
\end{equation}
where $\mathrm{N_{DI}},\mathrm{N_{TI}}$ and N are the numbers of doubly-ionized, triply-ionized and all events, respectively. Here, we mainly focus on non-sequential double ionization (NSDI) and on non-sequential triple ionization (NSTI) events. NSDI and NSTI involve an electron accelerating in the laser field and coming back to the core to transfer energy to bound electrons  via a recollision. This energy transfer can lead to the escape of two electrons (NSDI) or three electrons (NSTI). Electronic correlation, a fundamental interaction, underlies this field-assisted recollision \cite{Corkum_1994}.

 To identify a recollision in either one of the two models, we monitor the Coulomb potential between all pairs of  a quasi-free and a bound electron. We identify the maxima in the inter-electronic Coulomb potential energy as  function of time. We label the times when the inter-electronic distance is minimum as recollision times $\mathrm{t_{rec}}$. Also, we define the ionization time of electron i, $\mathrm{t^i_{ion}}$, to be the time when the compensated energy becomes positive and remains positive thereafter \cite{Leopold_1979}. We used the same definition for $\mathrm{t^i_{ion}}$ in all our previous studies, see for instance \cite{PhysRevA.78.023411,Emmanouilidou_2011}. The compensated energy is given by
\begin{align*}
\varepsilon^{\mathrm{comp}}_{\mathrm{i}}(\mathrm{t})= \mathrm{ \frac{\mathbf{\tilde{p}}^2_{\mathrm{i}} }{2\mathrm{m_i}} + \frac{Q_1 Q_i}{|\mathbf{r}_1-\mathbf{r}_i|}  + \mathrm{\sum_{\substack{\; \mathrm{j=2} \\ \mathrm{j} \neq \mathrm{i}}}^{\mathrm{N}}\mathrm{c_{i,j}(t)}\mathrm{V_{eff}(\zeta_j,|\mathbf{r}_{1}-\mathbf{r}_{i}|})}},
\end{align*}
\begin{equation*}
\varepsilon^{\mathrm{comp}}_{\mathrm{i}}(\mathrm{t})= \mathrm{ \frac{\mathbf{\tilde{p}}^2_{\mathrm{i}} }{2\mathrm{m_i}} + \frac{Q_1 Q_i}{|\mathbf{r}_1-\mathbf{r}_i|}  + V_{H,i} },
\end{equation*}
for the ECBB- and the H-model respectively.
Moreover, a TI or DI event is labelled as direct if the energy transferred from  a recolliding electron to  bound electrons  suffices for the simultaneous ionization, shortly after recollision, of three or two electrons. In Appendix \ref{AppendixPathways}, we outline the algorithm used to label an event as direct. Here, we label the remaining events as delayed TI and DI events. 

\subsection{DI and TI ionization probabilities}
We find that the DI and  TI probabilities are consistently larger for the H-model for both $\alpha=2,4$ compared to the ECBB-model. This is consistent with the different initial conditions the bound electrons have in the two models. The initial momenta of the bound electrons are higher in the H-model versus the ECBB-model, compare \fig{prob_distribution_t0_heis}(b) with \fig{prob_distribution_t0_eff}(b). Also, the repulsive Heisenberg potential reduces the attraction of each electron from the core resulting in higher ionization probabilities. Regarding the DI probability, for the H-model, we find that at intensities $2\times 10^{14}$ $\mathrm{W/cm^2}$, $4\times 10^{14}$ $\mathrm{W/cm^2}$, and $5\times 10^{14}$ $\mathrm{W/cm^2}$ and 20 fs pulse duration the DI probability is consistently higher for  $\alpha=4$ compared to $\alpha=2$. However, while at the two smallest intensities the difference in the DI probability is small for the two values of $\alpha$, at $5\times 10^{14}$ $\mathrm{W/cm^2}$ the DI probability is almost 71 \% higher for $\alpha=4$ compared to $\alpha=2$. Hence, the DI probability depends significantly on the value of  $\alpha$, a disadvantage of the H-model.  In what follows, we consider $\alpha=2$ for the H-model, unless otherwise stated, since this value allows the electrons to access a larger phase space. As we increase the intensity from $2\times 10^{14}\; \mathrm{W/cm^2}$ to $5\times 10^{14}\; \mathrm{W/cm^2}$, we find that the ratio of the DI probabilities between the two models, $\mathrm{P^{ECBB}_{DI} / P^{H}_{DI} }, $ decreases from   1.1  to 0.4. However, the ratio of the TI probabilities  $\mathrm{P^{ECBB}_{TI} / P^{H}_{TI} } $ increases from  0.03  to 0.2. Hence, for the intensities considered here, the DI and TI probabilities are smaller for the ECBB-model.

\subsection{Distribution of the sum of the electron momenta}\label{sec:sum_momenta}
\begin{figure}[H]
\centering
\includegraphics[width=\linewidth]{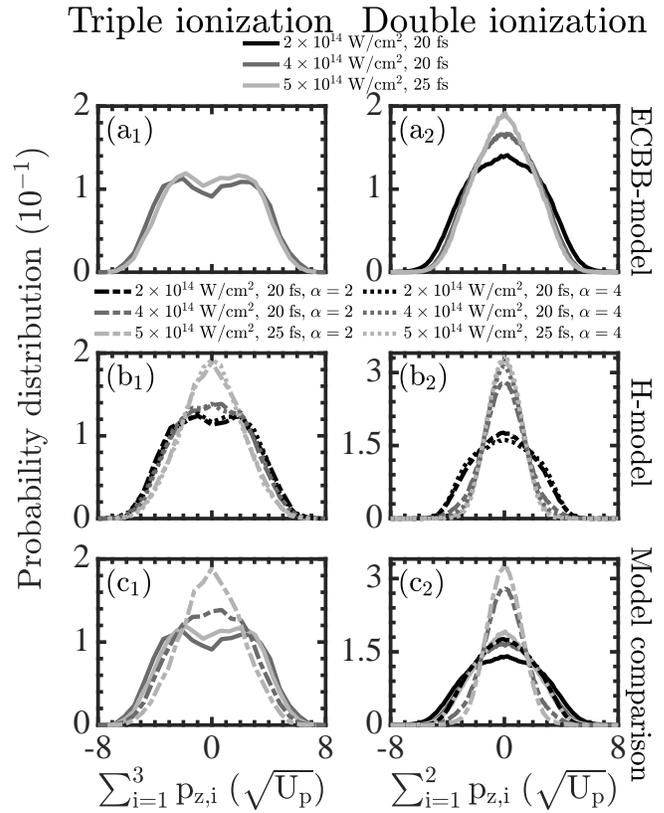}
\caption{Probability distributions of the sum of the  electron momentum components parallel to the polarization of the laser field for TI (left column) and DI (right column) at intensities $2 \times 10^{14} \; \mathrm{W/cm^2}$ (20 fs), $4 \times 10^{14} \; \mathrm{W/cm^2}$ (20 fs) and $5 \times 10^{14} \; \mathrm{W/cm^2}$ (25 fs). The ECBB-model results are presented in the top row, the H-model in the middle row and comparison of the two models in the third row. All probability distributions are normalized to one.}\label{fig:model_comparison}
\end{figure}

In \fig{fig:model_comparison}, we plot the TI and DI probability distribution of the sum of the $\mathrm{p_z}$ momenta of the ionizing electrons at intensities $2\times 10^{14}\; \mathrm{W/cm^2}$, $4\times 10^{14}\; \mathrm{W/cm^2}$ and 20 fs pulse duration  and $5\times 10^{14}\; \mathrm{W/cm^2}$ and 25 fs pulse duration. The  highest intensity pulse allows for a direct comparison with experimental results \cite{Rudenko_2008}.
In Figs. \ref{fig:model_comparison}(a1) and \ref{fig:model_comparison}(a2), for the ECBB-model, we plot the TI and DI probability distributions of the sum of the $\mathrm{p_z}$ momenta of the ionizing electrons. For DI (\fig{fig:model_comparison}(a2)), we find that the probability distribution is centered around zero and the width decreases  with increasing intensity. This is in accord with our previous findings for two-electron Ar driven by a 4 fs pulse  for intensities from $2 \times 10^{14} \; \mathrm{W/cm^2}$ to $5 \times 10^{14} \; \mathrm{W/cm^2}$\cite{ChenA2017Ndiw}. For TI (\fig{fig:model_comparison}(a1)), we find similar doubly-peaked distributions  for $4 \times 10^{14} \; \mathrm{W/cm^2}$ and $5 \times 10^{14} \; \mathrm{W/cm^2}$. The TI probability at $2 \times 10^{14} \; \mathrm{W/cm^2}$ is very low and we do not consider this intensity in \fig{fig:model_comparison}(a1).

In Figs. \ref{fig:model_comparison}(b1) and \ref{fig:model_comparison}(b2), for the H-model for $\alpha=2,4$, we plot the TI and DI probability distributions of the sum of the $\mathrm{p_z}$ momenta of the ionizing electrons. For DI (\fig{fig:model_comparison}(b2)), we find that the probability distribution is centered around zero and the width decreases with increasing intensity for both $\alpha$. For TI (\fig{fig:model_comparison}(b1)), we find that at $2 \times 10^{14} \; \mathrm{W/cm^2}$ the distribution is doubly-peaked, while  at  higher intensities the distribution is centered around zero. Finally, we find that the TI distributions are similar for the two values of $\alpha$, while the DI distributions are more centered around zero for $\alpha=4$. This is consistent with each electron being less attracted from the core for larger values of $\alpha$ resulting in smaller final momenta.  Hence, the probability distributions  depend on the value of  $\alpha$.

Comparing the TI  (\fig{fig:model_comparison}(c1)) and DI  (\fig{fig:model_comparison}(c2)) probability distributions of the ECBB- and  H-model for $\alpha=2$, we find that all distributions are more centered around zero for the H-model. Moreover,  the TI distributions at higher intensities are doubly-peaked for the ECBB-model and centered around zero for the H-model.

\begin{figure}[H]
\centering
\includegraphics[width=\linewidth]{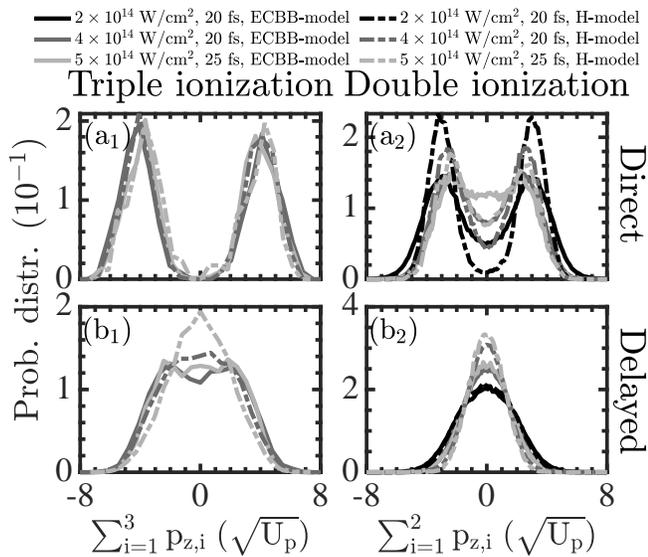}
\caption{Probability distributions of the sum of electron momenta components parallel to the polarization of the laser field for TI (left column) and DI (right column) at intensities  $2 \times 10^{14} \; \mathrm{W/cm^2}$ (20 fs), $4 \times 10^{14} \; \mathrm{W/cm^2}$ (20 fs) and $5 \times 10^{14} \; \mathrm{W/cm^2}$ (25 fs). The direct pathway distributions  are plotted on the top row and the delayed pathway distributions are plotted on the bottom row. All probability distributions are normalized to one.}\label{fig:model_comparison_mechanisms}
\end{figure}

In \fig{fig:model_comparison_mechanisms}, we plot, the TI and DI distributions of the sum of the p$_z$ electron momenta  for the direct (top row) and delayed pathway (bottom row). In the delayed pathway, we account for all non-direct events. Hence, here, the delayed events also include TI and DI no-recollision events. The latter  account  for roughly 7 \% of DI and 4 \% of TI events for the H- and  zero for the ECBB-model.

For the direct pathway, we find that the TI (\fig{fig:model_comparison_mechanisms}(a1)) and DI (\fig{fig:model_comparison_mechanisms}(a2)) distributions are double-peaked for both the ECBB- and the H-model. For TI events the  distributions have peaks at larger values of momenta   compared to DI events, with the peaks for TI being around $\pm 4$$\mathrm{\sqrt{U_p}}$ and for DI around $\pm 2.5$$\mathrm{\sqrt{U_p}}$. The ponderomotive energy $ \mathrm{U_p = E_0^2 / (4 \omega^2)} $ is the average energy an electron gains from the laser field. Also, for DI events,  the distributions have more events centered around zero for the ECBB- compared to the H-model. This contribution increases with increasing intensity, which is   consistent with our previous results  of double ionization of two-electron Ar driven by short pulses \cite{ChenA2017Ndiw}. 
 We find the percentage of direct events to be significantly larger for the ECBB- compared to the H-model. The contribution of direct events to DI is roughly 50 \% for the ECBB-model, while it decreases from 16 \% to 5 \% with increasing intensity for the H-model. The contribution of direct events to TI is roughly 20 \%  for the ECBB-model while it is roughly 5 \% for the H-model at the two highest intensities. For the delayed pathway, for DI, the distributions are centered around zero  for both models  (\fig{fig:model_comparison_mechanisms}(b2)), while for TI the distributions are less centered around zero  for the ECBB-model (\fig{fig:model_comparison_mechanisms}(b1)).

\begin{figure}[H]
\centering
\includegraphics[width=\linewidth]{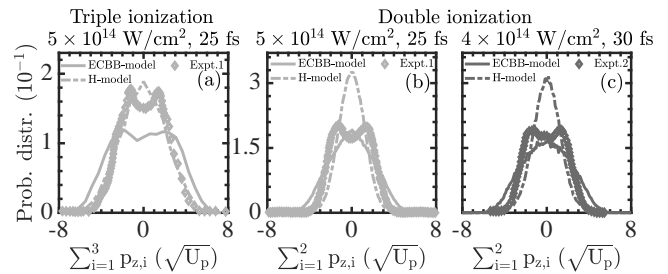}
\caption{Probability distributions of the sum of  electron momenta components parallel to the polarization of the laser field for TI (a) and DI (b) and (c) at intensities $5 \times 10^{14} \; \mathrm{W/cm^2}$ (25 fs) (a) and (b) and $4 \times 10^{14} \; \mathrm{W/cm^2}$ (30 fs) (c). We compare  the distributions obtained   with the ECBB- and H-model with experimental ones  \cite{Herrwerth_2008,Zrost_2006}. All probability distributions are normalized to one.}\label{fig:expt_comparison}
\end{figure}

 Next, we compare with experimental results  the findings of the ECBB- and H-model for the DI distribution  of the sum of the p$_z$  electron momenta of Ar at  $4 \times 10^{14} \; \mathrm{W/cm^2}$ (30 fs) (\fig{fig:expt_comparison}(c)) and $5 \times 10^{14} \; \mathrm{W/cm^2}$ (25 fs) (\fig{fig:expt_comparison}(b)) \cite{Herrwerth_2008,Zrost_2006} as well as the TI distribution at $5 \times 10^{14} \; \mathrm{W/cm^2}$ (25 fs) \cite{Zrost_2006} (\fig{fig:expt_comparison}(a)). The experimental DI distributions  have a slight double-peaked structure and agree more with the results of the ECBB-model. Indeed, the H-model produces DI distributions that are highly centered around zero,  which is significantly less the case for the ECBB-model. The experimental TI distributions  have a slight doubly-peaked structure which is only reproduced by the ECBB-model. However, the TI distribution obtained with the ECBB-model is wider than the one obtained experimentally. Hence, for DI the ECBB-model  better reproduces the experimental results while for TI is not clear whether the ECBB- or the H-model agree best with experiment. To answer this question a future study needs to compare distributions where intensity averaging has been accounted for in the theoretical results \cite{Wang:05,ChenA2017Ndiw}.

\subsection{Strength of the recollision in  DI and TI events}\label{sec:RecollisionStrength}
\begin{figure}[H]
\centering
\includegraphics[width=\linewidth]{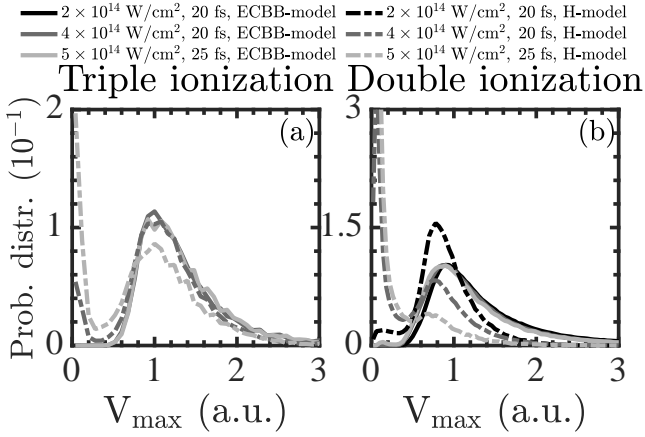}
\caption{Probability distributions of the largest value of the Coulomb inter-electronic potential energy  for TI (a) and DI (b) at intensities $2 \times 10^{14} \; \mathrm{W/cm^2}$ (20 fs), $4 \times 10^{14} \; \mathrm{W/cm^2}$ (20 fs) and $5 \times 10^{14} \; \mathrm{W/cm^2}$ (25 fs). All probability distributions are normalized to one.}\label{recollision_strength}
\end{figure}

For each DI and TI event we register all  the maxima of the Coulomb inter-electronic potential energy as a function of time and identify the largest maximum   V$_{\mathrm{max}}$. That is, we identify the most important recollision for each event. We plot the distribution of V$_{\mathrm{max}}$ for TI  (\fig{recollision_strength}(a)) and DI (\fig{recollision_strength}(b)) events. We find that recollisions are significantly stronger for the ECBB-model, with the most probable value of  V$_{\mathrm{max}}$ being roughly  1 a.u. 
for DI and TI at all intensities. In contrast, for the H-model, at the higher intensities,  the most probable value of  V$_{\mathrm{max}}$ is close to 0 a.u. both for DI and TI.  For the H-model, weaker recollisions are consistent with the DI and TI distributions of the sum of the p$_{z}$ electron momenta  being more centered around zero, see \fig{fig:model_comparison}.

\subsection{Correlated momenta}

\begin{figure}[H]
\centering
\includegraphics[width=0.8\linewidth]{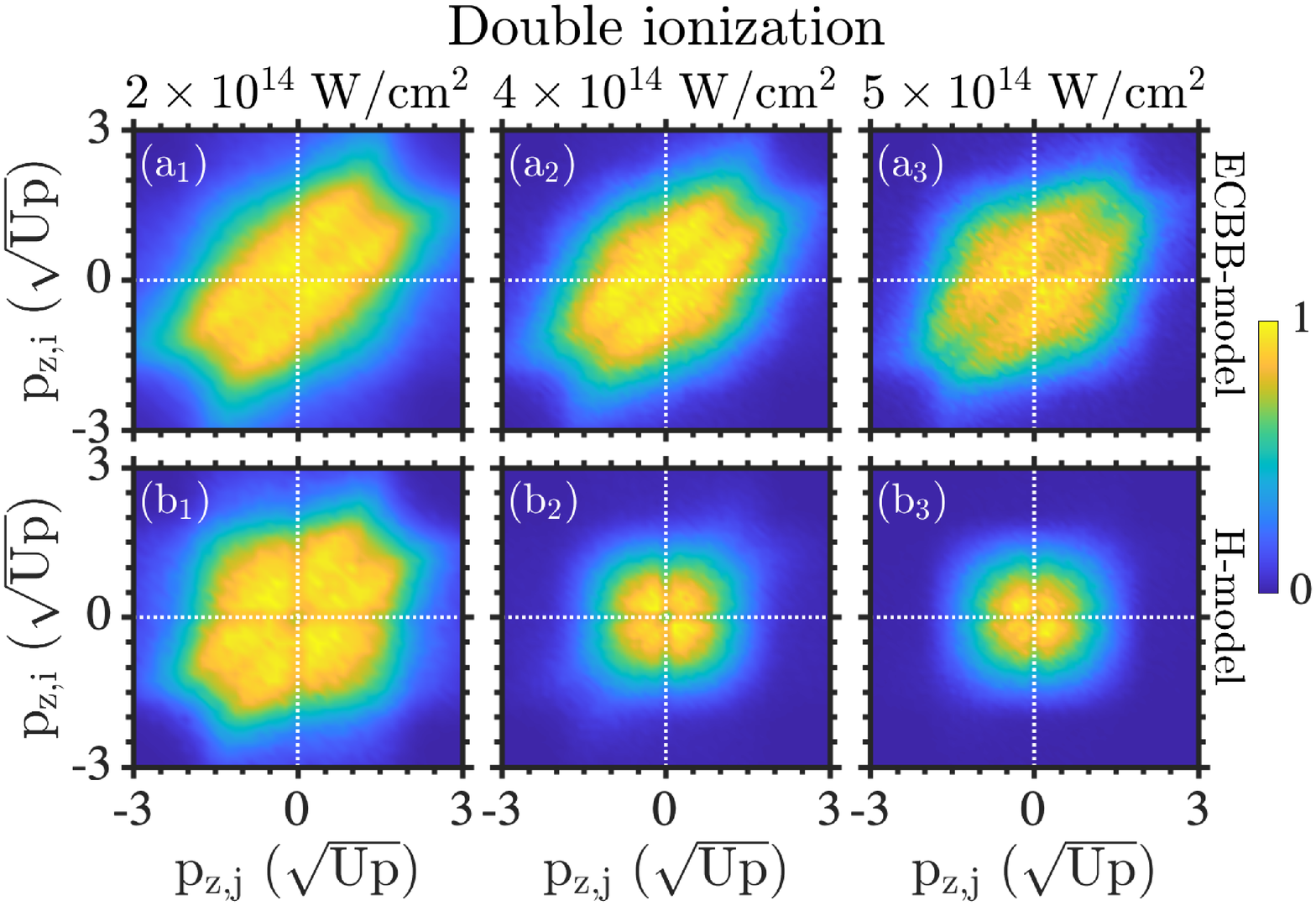}
\caption{Symmetrized correlated  momenta of all pairs of escaping electrons for DI for  the ECBB-model (a1)-(a3) and the H-model (b1)-(b3) at intensities $2 \times 10^{14} \; \mathrm{W/cm^2}$ (20 fs), $4 \times 10^{14} \; \mathrm{W/cm^2}$ (20 fs) and $5 \times 10^{14} \; \mathrm{W/cm^2}$ (25 fs).  The doubly differential distributions are divided by the peak value.}\label{correlated_momenta_20fs}
\end{figure}

In \fig{correlated_momenta_20fs}, for DI, we plot the correlated electron momenta   at intensities $2 \times 10^{14} \; \mathrm{W/cm^2}$ (20 fs), $4 \times 10^{14} \; \mathrm{W/cm^2}$ (20 fs) and $5 \times 10^{14} \; \mathrm{W/cm^2}$ (25 fs) obtained with the ECBB-model ((a1)-(a3))  and the H-model ((b1)-(b3)).  At the three intensities, we find that correlated electron escape prevails mostly for the  ECBB-model which produces roughly 10 \% more correlated events than the H-model.   Also, at intensities $4 \times 10^{14} \; \mathrm{W/cm^2}$  and $5 \times 10^{14} \; \mathrm{W/cm^2}$, the electrons escape with considerably higher momenta in the ECBB-model,  compare \fig{correlated_momenta_20fs} (a2) with \fig{correlated_momenta_20fs} (b2) and \fig{correlated_momenta_20fs} (a3) with \fig{correlated_momenta_20fs} (b3). The above are consistent with the ECBB-model resulting in more direct events (Sec. \ref{sec:sum_momenta}) and stronger recollisions (\fig{recollision_strength}) than the H-model.

\begin{figure}[H]
\centering
\includegraphics[width=0.8\linewidth]{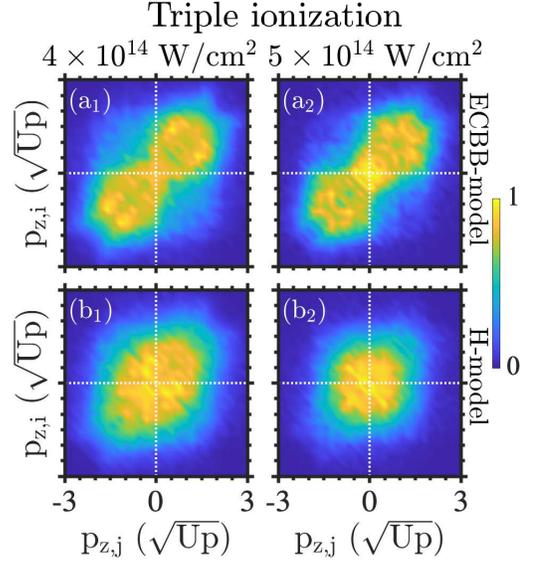}
\caption{Symmetrized correlated momenta of all  pairs of escaping electrons for TI for the ECBB-model (a1)-(a2) and the H-model (b1)-(b2)  at intensities $4 \times 10^{14} \; \mathrm{W/cm^2}$ (20 fs) and $5 \times 10^{14} \; \mathrm{W/cm^2}$ (25 fs). The doubly differential distributions  are divided by the peak value.}\label{correlated_momenta_20fs_TI}
\end{figure}

In \fig{correlated_momenta_20fs_TI}, for TI, we plot the  correlated electron momenta  at intensities $4 \times 10^{14} \; \mathrm{W/cm^2}$ (20 fs) and $5 \times 10^{14} \; \mathrm{W/cm^2}$ (25 fs) obtained with the ECBB-model  ((a1)-(a2)) and the H-model (b1)-(b2)). We find that correlated three electron-escape is clearly prevalent at both intensities for the ECBB-model, while this is barely the case for the H-model. Moreover, the three electrons escape with significantly smaller momenta for the H-model compared to the ECBB-model, compare \fig{correlated_momenta_20fs_TI} (a1) with \fig{correlated_momenta_20fs_TI} (b1) and \fig{correlated_momenta_20fs_TI} (a2) with \fig{correlated_momenta_20fs_TI} (b2). As for DI, this is consistent with the ECBB-model yielding  more direct events and stronger recollisions versus the H-model.

\subsection{Angular distributions}

\begin{figure}[H]
\centering
\includegraphics[width=\linewidth]{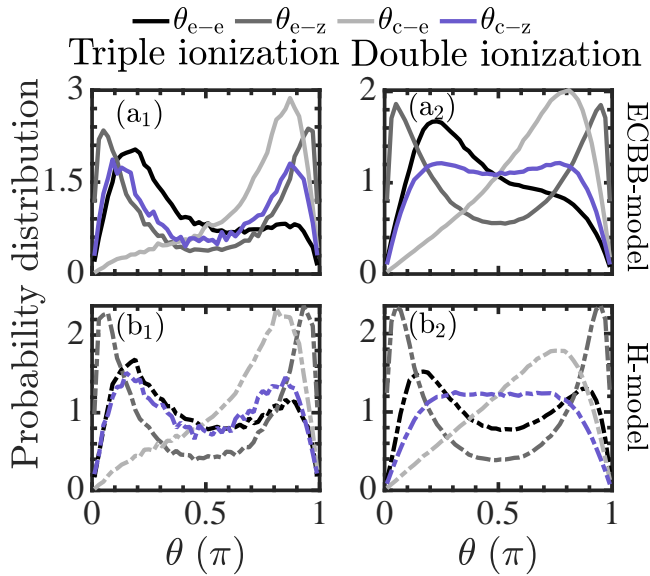}
\caption{  Probability distributions of the angles of the ionizing electrons and the core for TI (left column) and DI (right column)  at  $4 \times 10^{14} \; \mathrm{W/cm^2}$ (20 fs). Plots for the ECBB-model are denoted with solid lines versus broken lines for the  H-model. All probability distributions are normalized to one.}\label{angular_distributions}
\end{figure}

In \fig{angular_distributions}, we plot the TI (left column) and DI (right column) probability distributions of the angles of the ionizing electrons and the core  at intensity $4\times 10^{14}\; \mathrm{W/cm^2}$ at 20 fs. We obtain similar results for the other intensities considered in this work (not shown). We find  that the angle between any pair of escaping electrons $\theta_{e-e}$ (black color) is mostly peaked at small angles indicating a correlated electron escape.  For both the ECBB-model and the H-model, we find that the angle of inter-electronic escape is smaller for TI versus DI. This is consistent with the  electron momenta being more correlated for TI versus DI, compare \fig{correlated_momenta_20fs} with \fig{correlated_momenta_20fs_TI}. We also find that a small angle of inter-electronic escape is significantly more favoured by the ECBB-model, which results in more direct events and stronger recollisions.

We also find that the angle of any escaping electron with the z axis, $\mathrm{\theta_{e-z}}$, (dark gray color)  peaks at small and large values for TI and DI for both models. That is, the ionizing electrons escape mostly along (0$^\circ$) or opposite  (180$^\circ$) the direction of the electric field. However, the peaks of the distributions of $\mathrm{\theta_{e-z}}$ are sharper, i.e. the distributions are less wide, for the H-model. This is in accord with our finding that the H-model gives rise to a significantly higher number of TI and DI events where no-recollision takes place compared to the ECBB-model. As a result, in the H-model, the electrons ionize mostly due to the field for a larger number of events, with the electrons escaping more along or opposite the direction of the field.  Moreover, we find that the distributions of the angle of the core with the z axis, $\theta_{c-z}$, (blue color) are wide for both  TI and DI for both models. However, the distribution is wider for DI versus TI. This is consistent with the core having a higher charge equal to 3 for TI versus 2 for DI. As a result,  the electric field exerts a larger force  on the core  in TI  leading the core to escape more along or opposite the direction of the electric field. Finally, we find that the distribution of the angle $\theta_{c-e}$ (light grey color)
between an ionizing electron and the core peaks mostly at large angles, that is, the electron and the core escape in opposite directions. This is consistent with the electric field exerting opposite forces to particles of opposite charges. We find that the angle of escape between an electron and the core is larger for TI compared to DI for both models. This is consistent with the larger core charge for TI resulting to the core  escaping more along or opposite the field direction.

\section{Conclusions}
We formulate a 3D semi-classical model to address three-electron dynamics in a strongly driven atom where the electron and core dynamics are treated at the same time. Our formulation includes the magnetic field of the laser field  as well as  the Coulomb singularities. We address unphysical autoionization present in semi-classical models where the Coulomb singularities are accounted for and more than one electron is bound. We do so  by substituting the Coulomb repulsion between bound electrons with effective potentials where an effective charge is associated with every bound electron. The interaction between pairs of electrons that are not both bound is accounted for with the full Coulomb potential and all other forces are fully accounted for. This model, developed in this work and referred to as the ECBB-model,  identifies on the fly during time propagation if an electron is bound or not. We compare the ionization spectra obtained with the ECBB-model with the ones  obtained with a model previously developed--referred to here as the H-model. In the latter model,  a potential is added  for each electron that mimics the Heisenberg uncertainty principle and  restricts the accessible phase space of each electron   preventing autoionization. The advantage of the ECBB-model is that it accurately treats the interaction of each electron with the core and all other interactions while it treats less accurately the interaction between bound electrons. The advantage of the H-model is that it accurately treats the interaction between all electrons while it treats less accurately the interaction of each electron with the core.

Using these 3D semi-classical models, we address triple and double ionization in a strongly-driven atom, namely Ar. We compare the ionization spectra obtained with  the two models as well as with experiment for various pulse durations and intensities. We find that both double and triple ionization probabilities are greater for the H-model compared to the ECBB-model. We conjecture that this difference in the probabilities is due to the Heisenberg potential resulting in larger initial momenta of the bound electrons as well as in a significant less attraction of each electron from the core.  We find that in the H-model for a significant number of events the electrons ionize without a recollision, i.e. ionize due to the laser field and the recollisions are significantly weaker compared to the ECBB-model.  These findings are consistent with our results for the distribution of the sum of the momenta of the ionizing electrons along the direction of the laser field. For all the intensities and pulse durations considered here, we find that these distributions  are  broader for the ECBB-model.  For triple ionization, we find that  the distributions of the sum of the electron momenta have a double peak for  the ECBB-model while they are centered around zero for the H-model. We find  this to be due to the ECBB-model producing more direct ionization events than the H-model. This is also evident in the correlated electron momenta where the distributions obtained with the ECBB-model are consistently more correlated compared to the ones obtained with the H-model. Moreover, we identify another disadvantage of the H-model, namely, the distributions of the momenta and ionization probabilities depend on the parameter $\alpha$ in the Heisenberg potential.
Comparing with experimental distributions of the sum of the momenta  we find that the distributions obtained with the ECBB-model have a  better agreement with experiment mainly for double ionization. 
Finally, our formulation of the ECBB-model and of the H-model that account for electron and core motion and for non-dipole effects is general and can be generalized to strongly-driven atoms with more than three electrons.

\section{Acknowledgements}
A.E. and G.P.K. acknowledge the EPSRC Grant EP/W005352/1. The authors acknowledge the use of the UCL Myriad High Throughput Computing Facility (Myriad@UCL), and associated support services, in the completion of this work. Moreover, the authors are grateful to Prof. Armin Scrinzi for useful discussions.

\appendix
\section{Derivation of the effective potential}
\label{appendix:derivation}
The electric field produced by a charge $\mathrm{Q(\zeta_j,r)}$ that is contained within a spherical shell of radius r from the core is obtained by Gauss's law as follows: 
\begin{equation}\label{eq:A1}
\mathrm{\mathbf{E}(\zeta_j,r)  = \frac{Q(\zeta_j,r)}{ r^2}\hat{r}}.
\end{equation}
The work W done on a particle i due to the electric field $\mathrm{\mathbf{E}(\zeta_j,r)}$   is equal to minus the change in potential energy $\mathrm{\Delta V_{eff}} $:
\begin{align}\label{eq:work_1}
\begin{split}
\mathrm{W} &= \mathrm{-\Delta V_{eff}} \\
&= \mathrm{-\left[V_{eff}(\zeta_j,r) - V_{eff}(\zeta_j,\infty)\right]} \\
&= \mathrm{-V_{eff}(\zeta_j,r)}.
\end{split}
\end{align}
where we have used that $\mathrm{V_{eff}(\zeta_j,\infty) =0}.$ The work W is also given by
\begin{align}\label{eq:work_2}
\begin{split}
\mathrm{W} &= \mathrm{\int^r_{\infty}\mathbf{F} \cdot d\mathbf{r'}} \\
&= -\mathrm{\int^r_{\infty}\mathbf{E} \cdot d\mathbf{r'}} \\
&= -\mathrm{\int^r_{\infty}E(\zeta_j,r') \left(\hat{r}' \cdot \hat{r}'\right)dr' }\\
 &=  \mathrm{\int^r_{\infty} \left[  \frac{1}{r'^2}  - \frac{e^{-2\zeta_j r'} (1 + 2\zeta_j r')}{r'^2} - 2\zeta_j^2 e^{-2 \zeta_j r'} \right]dr' }\\
&=  \mathrm{\left(  -\frac{1}{r'}  + \frac{e^{-2\zeta_j r'} }{r'} +\zeta_j e^{-2 \zeta_j r'} \right)\Bigr|^r_{\infty} }\\
 &= -\mathrm{\frac{1 - (1+\zeta_j r)e^{-2\zeta_j r} }{r} },
\end{split}
\end{align}
where we have used that particle i is an electron and hence  $\mathrm{\mathbf{F}  = -\mathbf{E}  }$ as well as that particle j is an electron and $\mathrm{Q(\zeta_j,r)}$ is given by  Eq. (\ref{eqn::charge1}).  Using  Eqs. \eqref{eq:work_1} and \eqref{eq:work_2}, we find 
\begin{equation}
\mathrm{V_{eff}(\zeta_j,r)} = \mathrm{\frac{1 - (1+\zeta_j r)e^{-2\zeta_j r} }{r} },
\end{equation}
which is the potential energy that an electron i has at a distance r from the core due to a bound electron j. 

\section{Leapfrog Algorithm}\label{AppendixLeapfrog}
In what follows, we describe the leapfrog algorithm.  First, we initialize the auxiliary variables $\mathbf{W_{\mathrm{0}}^q}=\mathbf{q}_0,\mathbf{W^{\boldsymbol{\rho}}_{\mathrm{0}}}=\boldsymbol{\rho}_0,\mathrm{W_0^t}=\mathrm{t_0}$ and $\mathrm{W}^{\mathcal{E}}_0=\mathcal{E}_0.$ Then, we propagate for a  time step equal to $\mathrm{h}$, by propagating for half a step each quadruplet of variables ($\mathbf{q},\mathbf{W}^{\boldsymbol{\rho}}$,t,$\mathrm{W}^{\mathcal{E}}$) and ($\mathbf{W^q},\boldsymbol{\rho},\mathrm{W^t},\mathcal{E}$)  in an alternating way as follows
{\allowdisplaybreaks
\begin{align*}
\mathbf{q}_{1/2}&=\mathbf{q}_0+\dfrac{\mathrm{h}}{2}\dfrac{\dot{\mathbf{q}}(\mathbf{W_{\mathrm{0}}^q},\boldsymbol{\rho}_0,\mathrm{W_0^t})}{\Omega(\mathbf{W_{\mathrm{0}}^q})}\\
\mathbf{W}^{\boldsymbol{\rho}}_{1/2}&=\mathbf{W}^{\boldsymbol{\rho}}_{0}+\dfrac{\mathrm{h}}{2}\dfrac{\dot{\boldsymbol{\rho}}(\mathbf{W_{\mathrm{0}}^q},\boldsymbol{\rho}_0,\mathrm{W_0^t},\mathcal{E}_0)}{\Omega(\mathbf{W_{\mathrm{0}}^q})}\\
\mathrm{t}_{1/2}&=\mathrm{t}_{0}+\dfrac{\mathrm{h}}{2}\dfrac{1}{\Omega(\mathbf{W_{\mathrm{0}}^q})}\\
\mathrm{W}^{\mathcal{E}}_{1/2}&=\mathrm{W}^{\mathcal{E}}_{0}+\dfrac{\mathrm{h}}{2}\dfrac{\dot{\mathcal{E}}(\mathbf{W_{\mathrm{0}}^q},\boldsymbol{\rho}_0,\mathrm{W_0^t},\mathcal{E}_0)}{\Omega(\mathbf{W_{\mathrm{0}}^q})}\\
\mathbf{W_{\mathrm{1}}^q}&=\mathbf{W_{\mathrm{0}}^q}+\mathrm{h}\dfrac{\dot{\mathbf{q}}(\mathbf{q}_{1/2},\mathbf{W}^{\boldsymbol{\rho}}_{\mathrm{1/2}},\mathrm{t}_{1/2})}{\Omega(\mathbf{q}_{1/2})}\\
\boldsymbol{\rho}_{1}&=\boldsymbol{\rho}_{0}+\mathrm{h}\dfrac{\dot{\boldsymbol{\rho}}(\mathbf{q}_{1/2},\mathbf{W}^{\boldsymbol{\rho}}_{1/2},\mathrm{t}_{1/2},\mathrm{W}^{\mathcal{E}}_{1/2})}{\Omega(\mathbf{q}_{1/2})}\\
\mathrm{W_1^t}&=\mathrm{W_0^t}+\mathrm{h}\dfrac{1}{\Omega(\mathbf{q}_{1/2})}\\
\mathcal{E}_{1}&=\mathcal{E}_{0}+\mathrm{h}\dfrac{\dot{\mathcal{E}}(\mathbf{q}_{1/2},\mathbf{W}^{\boldsymbol{\rho}}_{1/2},\mathrm{t}_{1/2},\mathrm{W}^{\mathcal{E}}_{1/2})}{\Omega(\mathbf{q}_{1/2})}\\
\mathbf{q}_{1}&=\mathbf{q}_{1/2}+\dfrac{\mathrm{h}}{2}\dfrac{\dot{\mathbf{q}}(\mathbf{W_{\mathrm{1}}^q},\boldsymbol{\rho}_1,\mathrm{W_1^t})}{\Omega(\mathbf{W_{\mathrm{1}}^q})}\\
\mathbf{W}^{\boldsymbol{\rho}}_{1}&=\mathbf{W}^{\boldsymbol{\rho}}_{1/2}+\dfrac{\mathrm{h}}{2}\dfrac{\dot{\boldsymbol{\rho}}(\mathbf{W_{\mathrm{1}}^q},\boldsymbol{\rho}_{1},\mathrm{W_1^t},\mathcal{E}_1)}{\Omega(\mathbf{W_{\mathrm{1}}^q})}\\
\mathrm{t}_{1}&=\mathrm{t}_{1/2}+\dfrac{\mathrm{h}}{2}\dfrac{1}{\Omega(\mathbf{W_{\mathrm{1}}^q})}\\
\mathrm{W}^{\mathcal{E}}_{1}&=\mathrm{W}^{\mathcal{E}}_{1/2}+\dfrac{\mathrm{h}}{2}\dfrac{\dot{\mathcal{E}}(\mathbf{W_{\mathrm{1}}^q},\boldsymbol{\rho}_1,\mathrm{W_1^t},\mathcal{E}_1)}{\Omega(\mathbf{W_{\mathrm{1}}^q})}
\end{align*}}
\noindent The subscripts 0,1/2,1 denote the value of each variable at the start, the middle and the end of the time step h.

Next, we describe the algorithm that incorporates the leapfrog method in the Bulirsch-Stoer extrapolation scheme over a step H, which is split into n sub steps of size $\mathrm{h=H/n}$:

{\allowdisplaybreaks
\begin{align*}
\mathbf{q}_{1/2}&=\mathbf{q}_0+\dfrac{\mathrm{h}}{2}\dfrac{\dot{\mathbf{q}}(\mathbf{W_{\mathrm{0}}^q},\boldsymbol{\rho}_0,\mathrm{W_0^t})}{\Omega(\mathbf{W_{\mathrm{0}}^q})}\\
\mathbf{W}^{\boldsymbol{\rho}}_{1/2}&=\mathbf{W}^{\boldsymbol{\rho}}_{0}+\dfrac{\mathrm{h}}{2}\dfrac{\dot{\boldsymbol{\rho}}(\mathbf{W_{\mathrm{0}}^q},\boldsymbol{\rho}_0,\mathrm{W_0^t},\mathcal{E}_0)}{\Omega(\mathbf{W_{\mathrm{0}}^q})}\\
\mathrm{t}_{1/2}&=\mathrm{t}_{0}+\dfrac{\mathrm{h}}{2}\dfrac{1}{\Omega(\mathbf{W_{\mathrm{0}}^q})}\\
\mathrm{W}^{\mathcal{E}}_{1/2}&=\mathrm{W}^{\mathcal{E}}_{0}+\dfrac{\mathrm{h}}{2}\dfrac{\dot{\mathcal{E}}(\mathbf{W_{\mathrm{0}}^q},\boldsymbol{\rho}_0,\mathrm{W_0^t},\mathcal{E}_0)}{\Omega(\mathbf{W_{\mathrm{0}}^q})}\\
\mathbf{W_{\mathrm{1}}^q}&=\mathbf{W_{\mathrm{0}}^q}+\mathrm{h}\dfrac{\dot{\mathbf{q}}(\mathbf{q}_{1/2},\mathbf{W}^{\boldsymbol{\rho}}_{\mathrm{1/2}},\mathrm{t}_{1/2})}{\Omega(\mathbf{q}_{1/2})}\\
\boldsymbol{\rho}_{1}&=\boldsymbol{\rho}_{0}+\mathrm{h}\dfrac{\dot{\boldsymbol{\rho}}(\mathbf{q}_{1/2},\mathbf{W}^{\boldsymbol{\rho}}_{1/2},\mathrm{t}_{1/2},\mathrm{W}^{\mathcal{E}}_{1/2})}{\Omega(\mathbf{q}_{1/2})}\\
\mathrm{W_1^t}&=\mathrm{W_0^t}+\mathrm{h}\dfrac{1}{\Omega(\mathbf{q}_{1/2})}\\
\mathcal{E}_{1}&=\mathcal{E}_{0}+\mathrm{h}\dfrac{\dot{\mathcal{E}}(\mathbf{q}_{1/2},\mathbf{W}^{\boldsymbol{\rho}}_{1/2},\mathrm{t}_{1/2},\mathrm{W}^{\mathcal{E}}_{1/2})}{\Omega(\mathbf{q}_{1/2})}\\
\vdots\\
\mathbf{q}_{\mathrm{m}-1/2}&=\mathbf{q}_{\mathrm{m}-3/2}+\mathrm{h}\dfrac{\dot{\mathbf{q}}(\mathbf{W_{\mathrm{m-1}}^q},\boldsymbol{\rho}_{\mathrm{m}-1},\mathrm{W^t_{\mathrm{m}-1}})}{\Omega(\mathbf{W_{\mathrm{m-1}}^q)}}\\
\mathbf{W_{\mathrm{m-1/2}}^{\boldsymbol{\rho}}}&=\mathbf{W_{\mathrm{m-3/2}}^{\boldsymbol{\rho}}}+\mathrm{h}\dfrac{\dot{\boldsymbol{\rho}}(\mathbf{W^q_{\mathrm{m-1}}},\boldsymbol{\rho}_{\mathrm{m}-1},\mathrm{W^t_{\mathrm{m-1}}},\mathcal{E}_{\mathrm{m-1}})}{\Omega(\mathbf{W_{\mathrm{m-1}}^q)}}\\
\mathrm{t}_{\mathrm{m}-1/2}&=\mathrm{t}_{\mathrm{m}-3/2}+\mathrm{h}\dfrac{1}{\Omega(\mathbf{W_{\mathrm{m-1}}^q)}}\\
\mathrm{W}^{\mathcal{E}}_{\mathrm{m-1/2}}&=\mathrm{W}^{\mathcal{E}}_{\mathrm{m-3/2}}+\mathrm{h}\dfrac{\dot{\mathcal{E}}(\mathbf{W_{\mathrm{m-1}}^q},\boldsymbol{\rho}_{\mathrm{m-1}},\mathrm{W_{m-1}^t},\mathcal{E}_{\mathrm{m-1}})}{\Omega(\mathbf{W_{\mathrm{m-1}}^q})}\\
\mathbf{W^q_\mathrm{m}}&=\mathbf{W^q}_{\mathrm{m}-1}+\mathrm{h}\dfrac{\dot{\mathbf{q}}(\mathbf{q}_{\mathrm{m}-1/2},\mathbf{W_{\mathrm{m-1/2}}^{\boldsymbol{\rho}}},\mathrm{t}_{\mathrm{m}-1/2})}{\Omega(\mathbf{q}_{\mathrm{m}-1/2})}\\
\boldsymbol{\rho}_{\mathrm{m}}&=\boldsymbol{\rho}_{\mathrm{m}-1}+\mathrm{h}\dfrac{\dot{\boldsymbol{\rho}}(\mathbf{q}_{\mathrm{m}-1/2},\mathbf{W_{\mathrm{m-1/2}}^{\boldsymbol{\rho}}},\mathrm{t}_{\mathrm{m}-1/2},\mathrm{W}^{\mathcal{E}}_{\mathrm{m-1/2}})}{\Omega(\mathbf{q}_{\mathrm{m}-1/2})}\\
\mathrm{W_\mathrm{m}^t}&=\mathrm{W_{\mathrm{m}-1}^t}+\mathrm{h}\dfrac{1}{\Omega(\mathbf{q}_{\mathrm{m}-1/2})}\\
\mathcal{E}_{\mathrm{m}}&=\mathcal{E}_{\mathrm{m-1}}+\mathrm{h}\dfrac{\dot{\mathcal{E}}(\mathbf{q}_{\mathrm{m-1/2}},\mathbf{W}^{\boldsymbol{\rho}}_{\mathrm{m}-1/2},\mathrm{t_{m-1/2}},\mathrm{W}^{\mathcal{E}}_{\mathrm{m-1/2}})}{\Omega(\mathbf{q}_{\mathrm{m-1/2}})}\\
\vdots\\
\mathbf{W^q_\mathrm{n}}&=\mathbf{W^q}_{\mathrm{n}-1}+\mathrm{h}\dfrac{\dot{\mathbf{q}}(\mathbf{q}_{\mathrm{n}-1/2},\mathbf{W_{\mathrm{n-1/2}}^{\boldsymbol{\rho}}},\mathrm{t}_{\mathrm{n}-1/2})}{\Omega(\mathbf{q}_{\mathrm{n}-1/2})}\\
\boldsymbol{\rho}_{\mathrm{n}}&=\boldsymbol{\rho}_{\mathrm{n}-1}+\mathrm{h}\dfrac{\dot{\boldsymbol{\rho}}(\mathbf{q}_{\mathrm{n}-1/2},\mathbf{W_{\mathrm{n-1/2}}^{\boldsymbol{\rho}}},\mathrm{t}_{\mathrm{n}-1/2},\mathrm{W}^{\mathcal{E}}_{\mathrm{n-1/2}})}{\Omega(\mathbf{q}_{\mathrm{n}-1/2})}\\
\mathrm{W_\mathrm{n}^t}&=\mathrm{W_{\mathrm{n}-1}^t}+\mathrm{h}\dfrac{1}{\Omega(\mathbf{q}_{\mathrm{n}-1/2})}\\
\mathcal{E}_{\mathrm{n}}&=\mathcal{E}_{\mathrm{n-1}}+\mathrm{h}\dfrac{\dot{\mathcal{E}}(\mathbf{q}_{\mathrm{n-1/2}},\mathbf{W}^{\boldsymbol{\rho}}_{\mathrm{n}-1/2},\mathrm{t_{n-1/2}},\mathrm{W}^{\mathcal{E}}_{\mathrm{n-1/2}})}{\Omega(\mathbf{q}_{\mathrm{n-1/2}})}\\
\mathbf{q}_{\mathrm{n}}&=\mathbf{q}_{\mathrm{n}-1/2}+\dfrac{\mathrm{h}}{2}\dfrac{\dot{\mathbf{q}}(\mathbf{W^q_\mathrm{n}},\boldsymbol{\rho}_{\mathrm{n}},\mathrm{W^t_\mathrm{n}})}{\Omega(\mathbf{W^q_\mathrm{n}})}\\
\mathbf{W}^{\boldsymbol{\rho}}_{\mathrm{n}}&=\mathbf{W}^{\boldsymbol{\rho}}_{\mathrm{n-1/2}}+\dfrac{\mathrm{h}}{2}\dfrac{\dot{\boldsymbol{\rho}}(\mathbf{W_{\mathrm{n}}^q},\boldsymbol{\rho}_{\mathrm{n}},\mathrm{W_{\mathrm{n}}^t},\mathcal{E}_{\mathrm{n}})}{\Omega(\mathbf{W^q_\mathrm{n}})}\\
\mathrm{t}_{\mathrm{n}}&=\mathrm{t}_{\mathrm{n}-1/2}+\dfrac{\mathrm{h}}{2}\dfrac{1}{\Omega(\mathbf{W^q_\mathrm{n}})}\\
\mathrm{W}^{\mathcal{E}}_{\mathrm{n}}&=\mathrm{W}^{\mathcal{E}}_{\mathrm{n-1/2}}+\dfrac{\mathrm{h}}{2}\dfrac{\dot{\mathcal{E}}(\mathbf{W_{\mathrm{n}}^q},\boldsymbol{\rho}_{\mathrm{n}},\mathrm{W_{n}^t},\mathcal{E}_{\mathrm{n}})}{\Omega(\mathbf{W_{\mathrm{n}}^q})}
\end{align*}}
where m=2,...,n-1. 

\section{Identifying the direct pathway of TI and DI events}\label{AppendixPathways}
We obtain the TI and DI events with a code that incorporates the formulation of the ECBB-model described in Sec. \ref{effective_method} and a code that incorporates the formulation of the H-model described in Sec. \ref{heisenberg_method}. Once we obtain these events, we perform a detailed analysis with a different set of codes. In both analysis codes, i.e. one for each model, we use the framework we developed in Sec. \ref{switching} to determine on the fly during propagation if an  electron is quasi-free or bound.
 We register a TI or DI event as direct if a recollision is associated with the simultaneous ionization of three or two electrons.
We take the following   steps  to identify direct  events:
\begin{enumerate}
\item{We find the ionization time of each electron, $\mathrm{t^i_{ion}},$ with i = 1,2,3 for TI and i = 1,2 for DI.}
\item{ We register the maxima in the inter-electronic potential energies as a function of time between electron pairs i,j and  i,k and j,k during the time intervals when in these pairs one electron is quasi-free and the other is bound. Next, for each electron i, we identify the maximum for each one of the i,j and i,k potential energies that is closest to the time $\mathrm{t^i_{ion}}.$ We denote these times as $\mathrm{t_{rec}^{i,j}}$ and $\mathrm{t_{rec}^{i,k}}$. We obtain at  most six such times for TI events and four for DI events.}

\item{For each time $\mathrm{t_{rec}^{i,j}}$ we identify the time t$_2$ (see Sec. \ref{switching}) of closest approach to the core of the quasi-free electron (either electron i or j)  that is closest to $\mathrm{t_{rec}^{i,j}}$ and denote it as $\mathrm{t^{i,j}_2}$. We obtain at most six such times for TI events and four for DI events.}
\end{enumerate}
We label a TI event as direct if four of the times $\mathrm{t^{i,j}_2}$ are the same, accounting for one electron being  quasi-free and the other two bound. That is, if electron i is quasi-free during the recollision closest to the ionisation time $\mathrm{t^i_{ion}}$ then the times $\mathrm{t^{i,j}_2}$, $\mathrm{t^{i,k}_2}$, $\mathrm{t^{j,i}_2}$ and $\mathrm{t^{k,i}_2}$ should be the same. The times $\mathrm{t^{j,i}_2}$ and $\mathrm{t^{k,i}_2}$ are associated with the recollision times $\mathrm{t^{j,i}_{rec}}$ and $\mathrm{t^{k,i}_{rec}}$ for the bound electron j and k respectively. For the quasi-free electron we obtain two recollision times $\mathrm{t^{i,j}_{rec}}$ and $\mathrm{t^{i,k}_{rec}}$ associated with the ionization time $\mathrm{t^i_{ion}}$. We choose  the one that has the largest difference from  $\mathrm{t^i_{ion}}$, guaranteeing a stricter criterion for direct TI events. Next, we check whether $\mathrm{|t^{i,j}_{rec} - t^i_{ion}| < t_{diff}}$ or $(\mathrm{t^i_{ion} < t^{i,j}_{rec}} \; \& \; \mathrm{t^i_{ion} < t^{i,k}_{rec}})$ and $\mathrm{|t^{j,i}_{rec} - t^j_{ion}| < t_{diff}}$ and $\mathrm{|t^{k,i}_{rec} - t^k_{ion}| < t_{diff}}$. If the latter conditions are satisfied then we label the event as direct TI. The condition $(\mathrm{t^i_{ion} < t^{i,j}_{rec}} \; \& \; \mathrm{t^i_{ion} < t^{i,k}_{rec}})$ has also been used in our previous studies \cite{ChenA2017Ndiw,PhysRevLett.121.263203} to account for  a quasi-free electron ionising significantly earlier before recollision. This happens mostly at high intensities.  A similar process is followed to identify a DI event.

The interval $\mathrm{t_{diff}}$ is defined as the time duration where the inter-electronic potential energy undergoes a sharp change due to recollision. For the laser field intensities considered in this work, we find $\mathrm{t_{diff}}$ to be roughly equal to 1/8 laser cycle (T) for the ECBB-model and 1/6 T for the H-model. The difference in $\mathrm{t_{diff}}$ between the two models is due to the stronger electron-core interaction for the ECBB-model resulting in sharper changes to the electron-electron interaction. The choice of $\mathrm{t_{diff}}$ does not significantly change the percentage contribution of  direct TI and DI events \cite{ChenA2017Ndiw}.

\bibliography{bibliography}{}

\end{document}